\documentclass[%
reprint,
 amsmath,amssymb,
 aps,
]{revtex4-1}

\usepackage{graphicx}% Include figure files
\usepackage{dcolumn}% Align table columns on decimal point
\usepackage{bm}% bold math
\usepackage[normalem]{ulem} % underline 
\usepackage{color}

\usepackage{calc} % Calculate lengths using basic arithmetic operations
\newlength{\diff}
\newcommand{\bfe}[1]{\,{\bf e}_{\hat{#1}}}
\def\d{{\rm d}}
\def\e{{\rm e}}

\usepackage{soul}%\hl
\usepackage{xcolor}% gray

\begin{document}

% Length to separate lines in matrices with fractions:
\newlength{\matrixlinesep}
\setlength{\matrixlinesep}{0.2em}

\preprint{APS/123-QED}

\title{Visualizing gravitational Bessel waves}% Force line breaks with \\

\author{Ale\v{s} Pat\'{a}k}
\author{Martin Zouhar}%
\email{zouharm@isibrno.cz}
\author{Tom\'{a}\v{s} \v{R}ih\'{a}\v{c}ek}
\author{Oto Brzobohat\'{y}}
\affiliation{%
Institute of Scientific Instruments of the Czech Academy of Sciences, Kr\'{a}lovopolsk\'{a} 147, 612 64 Brno, Czech Republic
}%

\author{Jan Ger\v{s}l}
\author{Anna Charv\'{a}tov\'{a} Campbell}
\affiliation{
 Czech Metrology Institute, Okru\v{z}n\'{i} 31, 63800 Brno, Czech Republic
}%

       % but any date may be explicitly specified

\begin{abstract}
We explicitly derive a~vortex inspired solution for the metric perturbation within the linearized Einsteins general theory of relativity in arbitrary dimensions \mbox{$D\geq 4$}.
We focus on \mbox{$D=4$} where our solution is the gravitational analog of the well-known electromagnetic (or electron) Bessel vortex beams.
Next we visualize the perturbed spacetime via tidal tendexes and frame-drag vortexes.
We display and analyze mostly two-dimensional sections of the tendexes and the vortexes for different values of an angular momentum of the wave solution.
Corresponding geodesic deviation equations are solved and the results are visualized.
We show that the physically most important quadrupolelike case leads to a~wave with rotating polarization.
We discuss asymptotical features of the found solution.
We provide also several 3D plots of tendex lines.
One of them concerns a~special cylindrical-like case and we utilize the topological classification of singularities of the depicted line fields as an approach to characterize the radiation field.
\begin{description}
\item[PACS numbers]
04.25.Nx, 04.30.--w, 04.20.--q, 02.40.Pc
\end{description}
\end{abstract}

%\pacs{04.25.Nx, 04.30.--w, 04.20.--q, 02.40.Pc}% PACS, the Physics and Astronomy
\maketitle

%\tableofcontents

\section{\label{sec:intro}Introduction}

Gravitational waves were observed rather recently~\cite{ligo}.
The topic is challenging in both its experimental and theoretical parts.
Nevertheless, it opens new horizons for exploring astrophysical phenomena.
The first successful measurement detected the coalescence of two massive black holes.
This breakthrough discovery was followed by the coalescence of two neutron stars~\cite{ligo2}.
These new types of data combined with well-developed techniques for measuring electromagnetic radiation, neutrinos and cosmic rays have established so-called multimessenger astronomy.
Moreover, the multimessenger approach has been applied to test relativity models extending the Einstein theory of relativity, e.g., multidimensional theories~\cite{pardo}.
This stimulates a~development of new detectors and one may expect that one will be able to extract more features from the signals via the next generation detecting devices.
This is the reason to study more exotic waves in unconventional settings.

There is a~well-established analogy between waves in electromagnetism and in gravity~\cite{barnett, owen}.
This makes one ponder which electromagnetic phenomena have their counterparts in general relativity.

Bessel beams are a~promising example.
Diffraction, the spreading of propagating waves, is a~feature of a~majority of waves that appear in nature.
Almost 30 years ago, however, a~so-called ``nondiffracting" beam was discovered by Durnin in the optical domain \cite{DurninPRL87}. 
Its lateral field distribution can be described by a~Bessel function of the first kind, and
unlike most laser beams, which spread upon propagation, the transverse distribution of these Bessel beams remains constant.
As with a~plane wave, such idealized Bessel beams would be of infinite transverse extent and carry an infinite amount of energy and,
therefore, they can be generated experimentally over a~limited spatial range only.
Bessel beams are not limited to the optical domain but can be observed and utilized at various scales from electron beams~\cite{Grillo_PhysRevX_2014} at the nanoscale and electromagnetic waves~\cite{IndebetouwJOSAA89,LapointeOPTLASTECH92,McGloinCP05,BrzobohatyOE08} at the microscale, to acoustic waves~\cite{MITRI20081604,JIMENEZ2015245} at the macroscale.
Furthermore, they can serve as tractor beams~\cite{gorlach}.

Optical approximations to Bessel beams can be prepared in the laboratory either by focusing a~Gaussian beam with an axicon lens~\cite{BrzobohatyOE08} to generate a~Bessel-Gauss beam, or simply by placing a~narrow annular aperture in the far field~\cite{DurninPRL87,DurninJOSAA87}.

Vortex beams, which include Bessel beams, have attracted significant attention in the field of electron microscopy in the past decade.
Their properties and methods of creation have been under intense investigation since their theoretical prediction~\cite{bliokh_semiclassical_2007} and first experimental realization~\cite{uchida_generation_2010,verbeeck_production_2010}.
Vortex beams with high mode purity are generated using the holographic reconstruction principle, employing either amplitude~\cite{mcmorran_electron_2011} or phase~\cite{grillo_highly_2014} diffraction holograms, in the electron microscope.

Unlike their photon counterpart, electron vortex beams are composed of charged particles and therefore they possess an induced magnetic moment.
Both magnetic moment and intrinsic orbital angular momentum allow for coupling to materials.
This feature promises interesting applications, such as investigating magnetic chiral properties~\cite{lloyd_quantized_2012,schattschneider_is_2014,schachinger_emcd_2017}, probing and controlling plasmons~\cite{ugarte_controlling_2016,guzzinati_probing_2017}, determining chirality of various objects~\cite{juchtmans_using_2015}, manipulating nanoparticles~\cite{verbeeck_how_2013} or polarizing the spin of electrons~\cite{karimi_spin--orbital_2012}, though many of these utilizations are still at early development stages.

One can expect that there is also a~gravity counterpart to these Bessel beams, i.e., gravitational waves at the cosmological scale.
Cylindrical gravitational waves were considered already by Einstein and Rosen~\cite{einstein1937}.
The exact cylindrical gravitational wave solution, called an Einstein-Rosen wave, with Bessel functions of zeroth order is discussed also in Weber's monograph~\cite{weber}.
Furthermore, a~Bessel gravitational solution of the zeroth order in linearized gravity is mentioned in the classical book of Misner, Thorne, and Wheeler (MTW)~\cite{mtw}.
The lack of a~general order solution is a~motivation to consider Bessel gravitational waves.
Also, Bessel functions arise in numerical relativity; e.g., large order Bessel functions appear in the Fourier transform of gravitational waves from pulsars~\cite{chishtie}.

There is a~series of papers~\cite{owen, nichols, zhang, nichols2, zimmerman} that introduced a~method for visualizing spacetime curvature via tendexes and vortexes.
The series shows what powerful tools the concepts of tendexes and vortexes are by examining several types of solutions -- from weak fields, including plane waves, to Kerr black holes.

To our best knowledge, the Bessel gravitational wave problem is not fully analyzed and also visualized in the current literature.
A spinorial solution has been found for \mbox{$D=4$} \cite{birula}, but the metric has neither been presented nor examined.
We fill in this gap and, moreover, we obtain a~multidimensional linearized spacetime.
In this paper we study the Bessel solution of the wave equation for the metric perturbation in the Lorenz gauge.
We apply the concept of tendexes and vortexes and we visualize the Bessel gravitational wave for different values of parameters entering our solution.
In order to simplify the physical interpretation we also use the standard method of visualizing displacements of test particles.

The reminder of this paper is organized as follows:
The Bessel solution is derived in Sec.~\ref{sec:calculations}.
Details of the calculations, focused on general dimension \mbox{$D\geq4$}, are presented in the Appendix~\ref{sec:general.D.solution}.
In the following, we restrict ourselves to the case of \mbox{$D=4$}.
We briefly summarize the concept of tendexes and vortexes in Sec.~\ref{sec:visualization.tools} and continue by computing the tidal and the frame-drag field of the Bessel solution.
Section~\ref{sec:analysis} contains an analysis of different cases of the gravitational Bessel wave, including cylindrical waves, and we provide visualizations via their tendex and vortex lines.
The main part of our visualizations consists of a~series of two-dimensional plots of tendex and vortex lines in selected regions.
Moreover, several~3D~plots and a~geodesic deviation plot are provided in specific subcases of the gravitational wave.
We give some concluding remarks in Sec.~\ref{sec:conclusions}.

Throughout this paper we employ the following conventions.
We suppose that the spacetime is \mbox{$D$-dimensional}, \mbox{$D\geq 4$}.
We use the geometric unit system \mbox{$c=G=1$}.
The spacetime metric has a~mostly plus signature and we use the same conventions as in MTW~\cite{mtw}.
The range of the Greek indices is from zero to \mbox{$D-1$};
spatial coordinates are labeled by lowercase Latin indices running from $1$ to \mbox{$D-1$}.
When providing an expression in an orthonormal basis, we emphasize this using hats over the corresponding indices.

\section{\label{sec:calculations}The linearized field equations and their solutions}

We suppose that the spacetime metric $g$ in the weak field approximation differs only slightly from the Minkowski metric $\eta,$
\begin{equation}
g_{\mu\nu}=\eta_{\mu\nu} + h_{\mu\nu}, \quad |h_{\mu\nu}| \ll 1 .
\label{metric}\end{equation}

It is useful to employ a~coordinate system adapted to the symmetries of the investigated problem, in this case the cylindrical coordinate system.
The visualization will be performed in either the adapted system or in Cartesian coordinates.
We express the background metric in Minkowski/Cartesian coordinates
$$(t, x, y, z):\; 
\left.\eta_{\mu\nu}\right|_{\rm Mink.}=\mathrm{diag}(-1, 1, 1, 1)$$
and for comparison also in the cylindrical coordinates
$$(t,r,\varphi, z):\; 
\left.\eta_{\mu\nu}\right|_{\rm Cyl.}=\mathrm{diag}(-1, 1, r^2, 1).$$

The linearized Einstein field equations in vacuum reduce to the wave equations for the metric perturbation
\begin{equation}
\Box h_{\mu\nu} = 0,
\label{fielde}\end{equation} 
in the so-called Lorenz gauge (which is sometimes called the Hilbert, Einstein, De Donder or Fock gauge)
\begin{equation}
h^{\mu}_{\ \nu;\mu} - \frac{1}{2} h_{;\nu} = 0, 
\label{Lorenzg}\end{equation}
where $\Box$ denotes the d'Alembertian operator in curvilinear coordinates, the semicolon denotes the covariant derivative, \mbox{$h=h^{\mu}_{\ \mu}$} and we raise and lower indices with the Minkowski metric.

Our starting ansatz is
\begin{equation}
h_{\mu\nu} = \Re\{ A_{\mu\nu} {\rm J}_{\ell}(\kappa r) {\rm e}^{{\rm i}\phi} \},\ 
\phi = \ell\varphi + kz - \omega t,
\label{ansatz}\end{equation}
where $A_{\mu\nu}$ is a~matrix to be determined.
The above solution is described by the parameters~$k, \kappa, \ell$ and $\omega$.
The corresponding quantities in optics are $z$- and $r$-components of the wave vector ($k$ and $\kappa$), the $z$-component of the total angular momentum $\ell$ and the frequency $\omega$, respectively.
${\rm J}_{\ell}$~denotes Bessel function of the first kind and $\ell$'th order and $\Re$ stands for the real part.
This ansatz is inspired by linearizing the exact cylindrical wave solution from~\cite{weber}.

A relatively simple solution to Eqs.~\eqref{fielde} and~\eqref{Lorenzg} is determined in the cylindrical coordinate system by
\begin{equation}
A_{\mu\nu}=
\begin{bmatrix}
-\frac{a}{\kappa^2}(\omega^2+k^2) & 0 & 0 & 2\frac{a}{\kappa^2}\omega k\\[\matrixlinesep]
\strut 0 & a~& 0 & 0 \\[\matrixlinesep]
\strut 0 & 0 & ar^2 & 0\\[\matrixlinesep]
2\frac{a}{\kappa^2}\omega k & 0 & 0 & -\frac{a}{\kappa^2}(\omega^2+k^2)
\end{bmatrix},
\label{amplitudes}\end{equation}
where $a$ is a~constant and a~dispersion relation \mbox{$\omega^2=k^2+\kappa^2$} is enforced by the wave equations.
Let us call this solution the ``gravitational Bessel wave".

This solution can be extended to a~general dimension \mbox{$D\geq 4$}.
We employ a~split of the spatial hypersurface of dimension \mbox{$D-1$} into two parts.
The first part is a~\mbox{$B$-dimensional} submanifold (lowercase Latin indices from the beginning of the alphabet, e.g., $a$, $b$) parametrized as a~sphere,
hence the radius~$r$ and the angle~$\varphi.$
The remaining second-part is a~submanifold (lowercase Latin indices from the middle of the alphabet, e.g., $m$, $n$) parametrized with $z$-like coordinates.
The solution has a~similar form as that in Eq.~\eqref{ansatz}, the $A_{\mu\nu}$ is richer and the radial dependence through the Bessel function ${\rm J}_{\ell}(\kappa r)$ is generalized.
A detailed derivation, that applies to the \mbox{$D=4$} case as well, can be found in the Appendix~\ref{sec:general.D.solution}.
This derivation, performed in Minkowski coordinates, complements the above brief one utilizing curvilinear coordinates.

The amplitude matrix in Eq.~\eqref{amplitudes}, valid for \mbox{$D=4$}, extends for \mbox{$D\geq 4$} to
\begin{equation}\label{multi.amplitude}
\begin{split}
A_{00} &= 
\frac{\omega^2+(D-3)k^2}{(2-D)\omega k^2}\tilde{c} + \frac{D-B-2}{2-D}\tilde{b}
\,,\\%\label{multi.amplitude.00}\\
A_{0m} &= \tilde{c}\frac{k_m}{k^2}
\,,\\%\label{multi.amplitude.0m}\\
A_{ab} &= -\left[
\frac{1}{2-D}\frac{\kappa^2}{\omega k^2}\tilde{c} + \frac{D-B-2}{2-D}\tilde{b}
\right]\delta_{ab}
\,,\\%\label{multi.amplitude.ab}\\
A_{mn} &= \left[\frac{-\tilde{c}\omega}{k^2} + \tilde{b}\right]\frac{k_mk_n}{k^2}
\\%\nonumber\\
& - \frac{1}{2-D}\left[
\frac{\kappa^2}{\omega k^2}\tilde{c} - B\tilde{b}\right]\delta_{mn}
\,%\label{multi.amplitude.mn}
\end{split}
\end{equation}
where $\tilde{b}$ and $\tilde{c}$ are free parameters of the solution and \mbox{$k^2 = \delta_{mn}k^mk^n$}.
The dispersion relation is simply generalized with the multidimensional $k^2.$

A nondivergent \mbox{$D\geq4$} solution is obtained in the case of \mbox{$B=2$}, the radial dependence is the same as in the \mbox{$D=4$} ansatz Eq.~\eqref{ansatz}.
This nondivergent \mbox{$D\geq4$} solution reduces to the above \mbox{$D=4$} one in~Eq.~\eqref{amplitudes}, if we identify \mbox{$\tilde{c}=2\omega k^2 a/\kappa^2$}.

It can be shown that the \mbox{$D=4$} solution is equivalent to the one reported in the recent paper~\cite{birula}.
The authors of~\cite{birula} use the spinorial formalism~\cite{penrosec, penroseb} and they have the following spinorial components
(a~misprint in $\phi_{0111}$ is corrected, and we have aligned their notation to ours by the relabeling \mbox{$k_z=k$}, \mbox{$M=\ell$}, \mbox{$\rho=r$})
\begin{equation*}
\begin{bmatrix}
\phi_{0000}\\[\matrixlinesep]
\phi_{0001}\\[\matrixlinesep]
\phi_{0011}\\[\matrixlinesep]
\phi_{0111}\\[\matrixlinesep]
\phi_{1111}
\end{bmatrix} = {\rm e}^{{\rm i}\phi}
\begin{bmatrix}
(\omega +k)^4 {\rm e}^{-2{\rm i}\varphi} {\rm J}_{\ell-2}(\kappa r)
\\[\matrixlinesep]
{\rm i}(\omega +k)^3 \kappa {\rm e}^{-{\rm i}\varphi} {\rm J}_{\ell-1}(\kappa r)
\\[\matrixlinesep]
-(\omega +k)^2 \kappa^2 {\rm J}_{\ell}(\kappa r)\\[\matrixlinesep]
-{\rm i}(\omega +k)\kappa^3 {\rm e}^{{\rm i}\varphi} {\rm J}_{\ell+1}(\kappa r)
\\[\matrixlinesep]
\kappa^4 {\rm e}^{2{\rm i}\varphi} {\rm J}_{\ell+2}(\kappa r)
\end{bmatrix}.
\end{equation*}
This spinorial solution is written in Cartesian coordinates ($\varphi,\ r$ are treated as functions of Cartesian coordinates).
In fact, one can define the self-dual analog $G$ of the Riemann curvature tensor and express its components $G_{0i0j}$ using the spinorial components (for details see~\cite{birula}).
We have calculated $G_{0i0j}$ in cylindrical coordinates for both the spinorial-based solution~\cite{birula} and our solution~\eqref{ansatz} and~\eqref{amplitudes}, linearized to the first order in the metric perturbation.
The two sets of the components are the same apart from a~multiplicative constant.

If we set \mbox{$\ell=k=0$} (see also Sec.~\ref{sec:analysis.0.limits}), then our gravitational Bessel wave reduces to the cylindrical gravitational wave in Exercise 35.3 of MTW~\cite{mtw}.
Of~course, the cylindrical symmetry is broken
in the general case of the gravitational Bessel wave, where we have the so-called
``screw symmetry" instead, generated by the sum of a~translation and a~rotation; see Eq.~\eqref{kill.vector}.

In the general case, there is a~Killing vector field given by the following formula with constants $o,\, p,\, q$
%\begin{equation}
$$
\xi = o\frac{\partial}{\partial t} + p\frac{\partial}{\partial \varphi} + q\frac{\partial}{\partial z},\quad
p\ell + qk - o\omega = 0.
$$
%\end{equation}
It can be shown that the Killing vector is always spacelike except for the case of \mbox{$op\ell\neq 0$}.
An inequality relation between the free components of the Killing vector ensures that the vector becomes timelike for some finite real values of $r$.

The purely spatial part, \mbox{$o = 0$}, of the general Killing vector field describes the screw symmetry
\begin{equation}
\xi = -k \frac{\partial}{\partial \varphi} + \ell \frac{\partial}{\partial z}.
\label{kill.vector}
\end{equation}
We refer the reader to~\cite{ilderton} for more information on the analysis of screw-symmetric gravitational and electromagnetic waves and movements of testing particles subjected to these waves from the point of view of integrable systems.

\section{\label{sec:visualization.tools}
Tendexes and vortexes}

%We use definitions as introduced e.g., in~\cite{nichols}.
Let us summarize definitions introduced in the series of papers dedicated to visualizations of spacetime curvature, namely our Ref.~\cite{nichols}.
A {\it tendex} denotes a~collection of tendex lines which are integral curves of unit eigenvectors of the~tidal field~$\mathcal{E}$ (as defined below in Eq.~\eqref{tidal.general}).
The associated eigenvalue along a~single curve is called the line's ``tendicity".
We will color tendex lines by the sign of their tendicity.
Red (solid) lines have negative tendicity; an object, e.g., a~person, oriented along them is stretched.
The blue (dashed, in 2D plots only) lines have positive tendicity; an object oriented along them is squeezed.

Similarly, a~{\it vortex} describes a~bundle of 
vortex lines which are integral curves of unit eigenvectors of the frame-drag
field~$\mathcal{B}$ [see Eq.~\eqref{frame.drag.general}].
The eigenvalue associated with a~given line is called the ``vorticity" of this line.
The red (solid) lines have negative vorticity.
A person oriented along these lines will have their head rotated counterclockwise -- i.e., to the right -- with respect to their feet.
The blue (dashed, in 2D plots only) lines have positive vorticity
and they rotate the person's head clockwise.

The color intensity of the lines will represent the strength of the tendicity (vorticity).
We will appropriately scale the color intensities for best visual presentations in our figures; the scaling will be achieved via the multiplicative factor and/or a~suitable choice of a~color map. 

One can split the Weyl curvature tensor~$C$ %$C_{\mu\nu\rho\sigma}$
into its so-called ``electric" part~$\mathcal{E}$, which can be interpreted as a~tidal field, and ``magnetic" part~$\mathcal{B}$, that can be interpreted as a~frame-drag field.
\begin{eqnarray}
\mathcal{E}_{\alpha\beta} &=& C_{\alpha\mu\beta\nu} e_{\hat{0}}^{\ \mu} e_{\hat{0}}^{\ \nu},\ \text{i.e.},\ 
\mathcal{E}_{ij}=C_{i\hat{0}j\hat{0}}.
\label{tidal.general}\\
\mathcal{B}_{\alpha\beta} &=& -\ast\! C_{\alpha\mu\beta\nu} e_{\hat{0}}^{\ \mu} e_{\hat{0}}^{\ \nu},\ \text{i.e.},\ 
\mathcal{B}_{ij}=\frac{1}{2}\epsilon_{ipq} C^{pq}_{\ \ j\hat{0}}.
\label{frame.drag.general}
\end{eqnarray}
The above equations use the following notation.
The symbol $\ast$ denotes the Hodge dual, 
$\ast C_{\rho\mu\sigma\nu} = \frac{1}{2}\epsilon_{\rho\mu\eta\lambda} C^{\eta\lambda}_{\ \ \sigma\nu}.$
Expressions with Latin (spatial) indices are written in \mbox{3+1} notation,
$e_{\hat{0}}^{\ \mu}$ is chosen to be part of an orthonormal tetrad and denotes the components of the foliation's unit time basis vector used in \mbox{3+1} split of the spacetime.
Furthermore, we use the MTW convention for the Levi-Civita tensor~\mbox{$\epsilon_{\hat{0}\hat{1}\hat{2}\hat{3}}=1$}, \mbox{$\epsilon_{ipq}=\epsilon_{\hat{0}ipq}$} and 
$\epsilon_{\hat{1}\hat{2}\hat{3}}=1$ in right-handed orthonormal frames.
Let us note that in vacuum, as is our case, the Weyl curvature tensor equals the Riemann tensor.

We will solve the eigenvalue problem for both tensors $\mathcal{E}$ and $\mathcal{B}$ in the orthonormal basis,
$$\mathcal{E}_{\hat{a}\hat{b}}v^{\hat{b}}= \lambda v_{\hat{a}},\
\mathcal{B}_{\hat{a}\hat{b}}v^{\hat{b}}= \lambda v_{\hat{a}},$$
and integration of the differential equation
%$$\frac{\d x^{\mu}}{\d s} = v^{\hat{a}}e_{\hat{a}}^{\ \mu}$$
$$\frac{\d x^{\mu}(s)}{\d s} = v^{\hat{a}}(x^{\rho}(s))\,e_{\hat{a}}^{\ \mu}(x^{\rho}(s))$$
%$$\frac{\d x^{\mu}(s)}{\d s} = v^{\hat{a}}\left(x^{\rho}(s)\right)e_{\hat{a}}^{\ \mu}\left(x^{\rho}(s)\right)$$
will follow to find their streamlines in a~coordinate basis ($s$ denotes a~parameter along the streamlines).

To accomplish this goal we have used Maple %\textsuperscript{\tiny TM}
to verify some symbolic calculations %(by hand)
and custom PYTHON scripts utilizing several libraries~\cite{numpy.article,numpy.man,scipy,sympy,matplotlib,ramachandran2011mayavi} to create visualizations of streamlines.
%To accomplish this goal we have performed symbolic calculations and applied custom python scripts to create streamlines visualizations.
These scripts use integration methods based on a~Runge-Kutta algorithm.
The standard numerical integration on a~grid was modified in order to cover for possible sign differences of eigenvectors at different grid points.
We had to decide which of the to-be-interpolated neighboring eigenvectors are to be flipped (change of sign); the criterion was based on the value of the dot product.
This complication reflects the fact that we are dealing with the so-called line field which
is not just a~vector field but its projective equivalent~\cite{crowley}.
In order to cover the whole range of displayed coordinates ``nicely", with streamlines extending up to the displayed boundaries and with a~reasonable density, we have used a~larger range and then we have removed the boundary regions.

In the case of the Minkowski spacetime perturbed by the gravitational Bessel wave [Eqs.~\eqref{ansatz} and~\eqref{amplitudes}], the general formulas from Eqs.~\eqref{tidal.general} and~\eqref{frame.drag.general} become 
\begin{widetext}
\begin{equation}
\mathcal{E}_{\hat{a}\hat{b}} =\frac{a}{2} 
\begin{bmatrix}
\left\{\left[E_1 - k^2\right]{\rm J}_{\ell} + E_0{\rm J}_{\ell+1}\right\}\cos\phi &
\left\{-E_1{\rm J}_{\ell} + E_0\ell {\rm J}_{\ell+1}\right\}\sin\phi &
\kappa k\left[\frac{\ell}{\kappa r}{\rm J}_{\ell} - {\rm J}_{\ell+1}\right]\sin\phi \\[\matrixlinesep]
\ast &
\left\{\left[-E_1 + \omega^2\right]{\rm J}_{\ell} - E_0{\rm J}_{\ell+1}\right\}\cos\phi &
\frac{kl}{r}{\rm J}_{\ell}\cos\phi \\[\matrixlinesep]
\ast &
\ast &
- \kappa^2{\rm J}_{\ell}\cos\phi \\[\matrixlinesep]
\end{bmatrix}
\label{tendexes.bessel}
,\end{equation}
\begin{equation}
\mathcal{B}_{\hat{a}\hat{b}} = \omega a
\begin{bmatrix}
\left\{B_1{\rm J}_{\ell} - B_0\ell {\rm J}_{\ell+1}\right\}k\sin\phi &
\left\{\left[B_1 - \frac{1}{2}\right]{\rm J}_{\ell} + B_0{\rm J}_{\ell+1}\right\}k\cos\phi &
-\frac{1}{2}B_0\ell {\rm J}_{\ell}\kappa\cos\phi
\\[\matrixlinesep]
\ast &
-\left\{B_1{\rm J}_{\ell} - B_0\ell {\rm J}_{\ell+1}\right\}k\sin\phi &
\frac{1}{2}\left\{B_0\ell {\rm J}_{\ell}-{\rm J}_{\ell+1}\right\}\kappa\sin\phi
\\[\matrixlinesep]
\ast &
\ast &
0
\\[\matrixlinesep]
\end{bmatrix}
\label{vortexes.bessel}
,\end{equation}
\end{widetext}
%\begin{equation*}
%\mathcal{E}_{\hat{a}\hat{b}} =\frac{a}{2} 
%\begin{bmatrix}
%\ast &
%\left(-\mathcal{J}_{\ell} + \ell\mathcal{J}_{\ell+1}\right)\sin\phi &
% \frac{\ell {\rm J}_{\ell}- \kappa r {\rm J}_{\ell+1}}{r} k\sin\phi
%\\
%\ast &
%-\left(\mathcal{J}_{\ell} - \omega^2 {\rm J}_{\ell} + \mathcal{J}_{\ell+1}\right)\cos\phi &
%\frac{kl}{r}{\rm J}_{\ell}\cos\phi
%\\
%\ast &
%\ast &
%- \kappa^2{\rm J}_{\ell}\cos\phi
%\\
%\end{bmatrix}
%\label{tendexes.bessel}
%\end{equation*}
%and
%\begin{equation*}
%\mathcal{B}_{\hat{a}\hat{b}} = k\omega a
%\begin{bmatrix}
%\ast &
%\left(\mathcal{J}_{\ell}^1 - \frac{1}{2}{\rm J}_{\ell} + \mathcal{J}_{\ell+1}^0\right)\cos\phi &
%-\frac{\kappa}{2k}\ell\mathcal{J}_{\ell}^0\cos\phi
%\\
%\ast &
%-\left(\mathcal{J}_{\ell}^1 - \ell\mathcal{J}_{\ell+1}^0\right)\sin\phi &
%\frac{\kappa\left(\mathcal{J}_{\ell}^0-{\rm J}_{\ell+1}\right)}{2k}\sin\phi
%\\
%\ast &
%\ast &
%0
%\\
%\end{bmatrix}
%\label{vortexes.bessel}
%\end{equation*}
where we have introduced some useful abbreviations
%${\rm J}_{\ell}\equiv {\rm J}_{\ell}(\kappa r)$, \mbox{$\mathcal{J}_{\ell} = E_1 {\rm J}_{\ell}$}, \mbox{$\mathcal{J}_{\ell+1} = E_0 {\rm J}_{\ell+1}$}, \mbox{$\mathcal{J}_{\ell}^m = B_m {\rm J}_{\ell}$}, with
$$E_m = \frac{\omega^2+k^2}{(\kappa r)^{m+1}}[\ell(\ell-1)]^m ,\ B_m = \frac{E_m}{\omega^2+k^2},\  {\rm J}_{\ell}\equiv {\rm J}_{\ell}(\kappa r).$$
The components indicated by~$\ast$ are determined by the symmetry of the matrix.
The tetrad is given by
\begin{equation}
\bfe{0}=\partial_t,\ \bfe{1}=\partial_r,\ \bfe{2}=\frac{1}{r} \partial_{\varphi}, 
\bfe{3}=\partial_z.
\label{ourtetrad}
\end{equation}
This orthonormal basis is of zeroth order in the metric perturbation which is sufficient 
because we work in the linearized theory.

\section{\label{sec:analysis}Visualization and discussion of the results}
%\section{\label{sec:analysis}Visualization of the results}
The calculated tidal and frame-drag fields presented in %Section~\ref{sec:visualization.bessel}
Eqs.~\eqref{tendexes.bessel} and~\eqref{vortexes.bessel}
are rather complicated because of a~large amount of nonzero components.
The case \mbox{$\ell=0$} is simple enough to be solved exactly.
For $\ell > 0$ we split the analysis into two limiting ranges of the coordinate $r$, namely $r~\approx~0$ and large values of $r$, and use only the leading terms in $r$ and \mbox{$1/r$} respectively.
This allows us to approximate the tidal and frame-drag fields but still also to visualize and to analyze the dominant features in these regions.
For the sake of brevity we do not explicitly write the coordinate dependencies of the leading terms of given approximations but we only express the approximated tendexes, vortexes and associated eigenvalues
using these leading components.

\subsection{\label{sec:analysis.l0.r}The case of \mbox{$\ell=0$}, arbitrary value of $r$}
A simple general solution of tendexes of the tidal field Eqs.~\eqref{tendexes.bessel} can be found in the case of \mbox{$\ell=0$}.
$$\lambda_{1, 2} = \frac{1}{2}\left[
-\mathcal{E}_{\hat{2}\hat{2}}\pm\sqrt{(\mathcal{E}_{\hat{1}\hat{1}}-\mathcal{E}_{\hat{3}\hat{3}})^2 + 4\mathcal{E}^2_{\hat{1}\hat{3}}}\right],\;
\lambda_3 = \mathcal{E}_{\hat{2}\hat{2}}$$
and the corresponding eigenvectors
\begin{equation}
{\bf v}_{1, 2} = \frac{-\mathcal{E}_{\hat{1}\hat{3}}\bfe{1}+(\mathcal{E}_{\hat{1}\hat{1}}-\lambda_{1, 2})\bfe{3}}{\sqrt{\mathcal{E}_{\hat{1}\hat{3}}^2+(\mathcal{E}_{\hat{1}\hat{1}}-\lambda_{1, 2})^2}},\;
{\bf v}_3 = \bfe{2}.
\label{vvt0}
\end{equation}
Similarly, the vortexes are given by
$$\lambda_{1, 2} = \pm\sqrt{\mathcal{B}_{\hat{1}\hat{2}}^2 + \mathcal{B}_{\hat{2}\hat{3}}^2},\;
\lambda_3 = 0$$
and
\begin{eqnarray}
\sqrt{2}{\bf v}_{1, 2} &=& \frac{\mathcal{B}_{\hat{1}\hat{2}}}{\lambda_{1, 2}}\bfe{1} + \bfe{2} + \frac{\mathcal{B}_{\hat{2}\hat{3}}}{\lambda_{1, 2}}\bfe{3},\label{vvv01}\\
{\bf v}_3 &=& \frac{-\mathcal{B}_{\hat{2}\hat{3}}\bfe{1}+\mathcal{B}_{\hat{1}\hat{2}}\bfe{3}}{\left|\lambda_{1, 2}\right|}.
\label{vvv02}
\end{eqnarray}

\begin{figure}[t]
\includegraphics[width=0.48\textwidth]{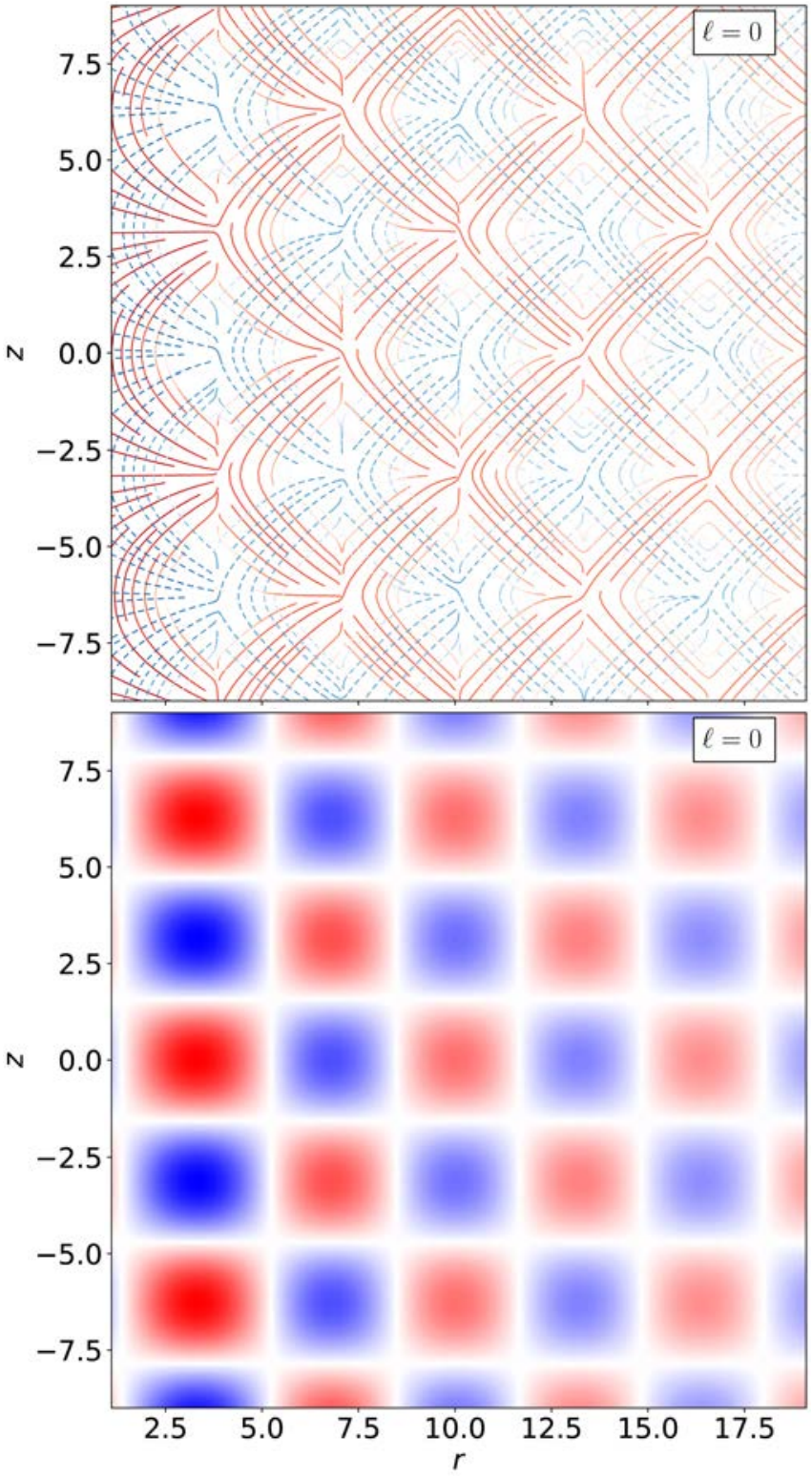}
\caption{\label{monot}
(color online).
Top: Tendexes in the case of \mbox{$\ell=0$} corresponding to the eigenvectors in the $(r, z)$ plane.
The tendicities are positive for dashed (blue) lines and negative for solid (red) lines.
Values of the other parameters are \mbox{$t=0$}, \mbox{$\varphi=0$} and \mbox{$k=\kappa=1$}.\\
Bottom: 
Tendicities in the case of \mbox{$\ell=0$} corresponding to the tendexes of $\varphi$-aligned eigenvector~${\bf v}_3$, see Eq.~\eqref{vvt0}, in the $(r, z)$ plane.
The other parameters have the same values as in the top panel.
The signs of the eigenvalues (colors) interchange in a~chessboard pattern.
The red regions have negative tendicity and the blue regions have positive tendicity.
The color intensity represents the magnitude of the tendicity.
The color at $(3.5, 6.7)$ is red (lighter gray).
}
\end{figure}

%Fig.~\ref{monot} displays a~quasi-periodic pattern of the tendexes in the $(r, z)$ plane.
Both panels in Fig.~\ref{monot} display a~quasiperiodic pattern of the tendexes in the $(r, z)$ plane.
The period in the~$r$ direction 
is not constant but it undergoes small changes as $r$ increases; this is due to the specifics of the tendexes' solution and it is also related to the roots of the Bessel functions.
The figure contains areas where tendexes with opposite signs of tendicity cross each other, which is a~common and expected feature.
Quite surprisingly, there are also places where one of the tendicities is zero in this $(r, z)$ section.
Part of the tendicity in the $(r, z)$ plane
must be ``transferred" to the $\varphi$-direction orthogonal to the displayed $(r, z)$ plane.
This is a~consequence of the fact that the tidal field~$\mathcal{E}$ is trace free which means the sum of its eigenvalues has to vanish.
This becomes clear when comparing the top plot and the bottom one in Fig.~\ref{monot} -- the total sum of their tendicities vanishes (roughly speaking, the blue cancels out the red).
It is also interesting that there are regions where tendicities in all three directions are nonzero.
This means that the general gravitational Bessel waves are not purely transversal -- unlike the plane gravitational wave -- in the region displayed.
These findings are also valid asymptotically for large $r$ and for all $\ell$; see Sec.~\ref{sec:analysis.l.r.large}.

\begin{figure}[h]
\includegraphics[width=0.48\textwidth]{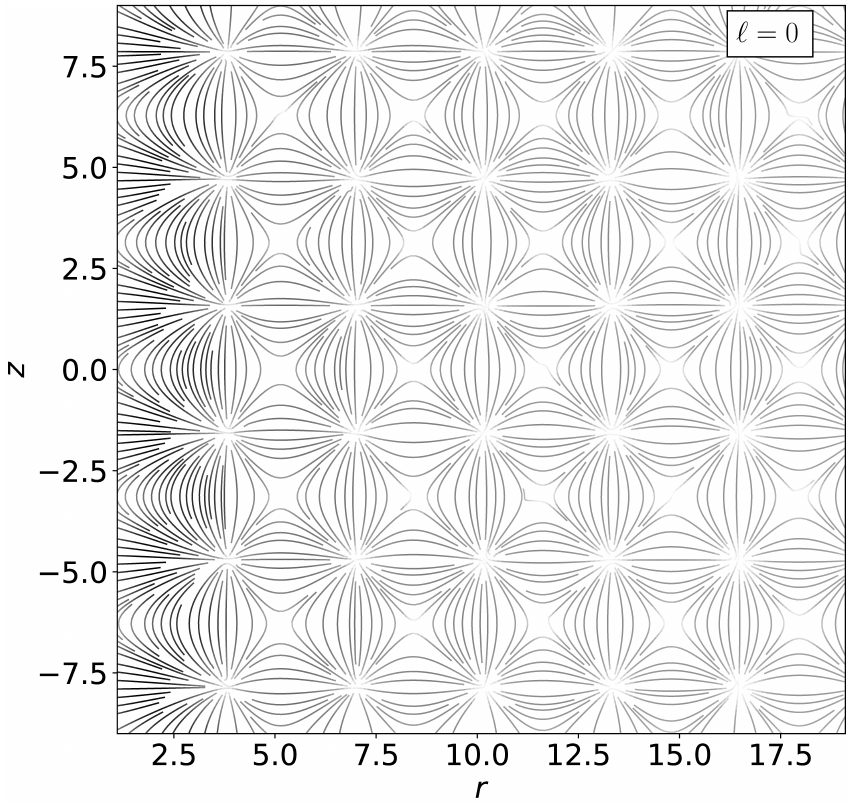}
\caption{\label{monov}
Vortex lines in the case of \mbox{$\ell=0$} in the $(r, z)$ plane corresponding to the case presented in~Fig.~\ref{monot}.
Vorticity of these lines is identically zero.
Shading of the lines is determined by the absolute value of the vorticity of the other two vortex lines. The other lines do not lie in the plane and have equal and opposite vorticities.
}
\end{figure}

The structure of vortex lines in the $(r, z)$ plane is indicated in~Fig.~\ref{monov}.
Equation~\eqref{vvv01} implies that the displayed $(r, z)$ section is a~plane of reflection
symmetry for the vortexes.
Indeed, flipping the sign of the $\bfe{2}$ component of ${\bf v}_1$ gives $-{\bf v}_2$ and the minus sign is irrelevant for eigenvectors and vortex analysis.
As a~result, the vortex lines corresponding to ${\bf v}_{1, 2}$ in Eq.~\eqref{vvv01} must cross the plane at equal and opposite inclinations and they have
equal and opposite vorticities. Figure~\ref{monov} displays the vortex lines with zero
vorticity, so they are not physically relevant in the sense to show a~differential frame dragging
along them. However, they display the orientation of the other two vortexes whose projection
onto the plane is orthogonal to the vortex lines shown. The shading of the displayed lines
represents the vorticities of the other two vortex lines.

\subsection{\label{sec:analysis.l1.r0}The case of \mbox{$\ell=1$}, small values of $r$}
The calculated tidal field, see Eq.~\eqref{tendexes.bessel}, simplifies so that one can approximate
$$\lambda_{1, 2} = \pm\sqrt{\mathcal{E}^2_{\hat{1}\hat{3}}+\mathcal{E}^2_{\hat{2}\hat{3}}},\;
\lambda_3 = 0$$
and
\begin{eqnarray}
\sqrt{2}{\bf v}_{1, 2} &=& \frac{\mathcal{E}_{\hat{1}\hat{3}}}{\lambda_{1, 2}}\bfe{1} + \frac{\mathcal{E}_{\hat{2}\hat{3}}}{\lambda_{1, 2}}\bfe{2} + \bfe{3}
,\label{vvt11}\\
{\bf v}_3 &=& \frac{-\mathcal{E}_{\hat{2}\hat{3}}\bfe{1} + \mathcal{E}_{\hat{1}\hat{3}}\bfe{2}}{\left|\lambda_{1, 2}\right|}
\label{vvt12}\end{eqnarray}
in the region of interest.
Similarly, vortexes are given by the same formulas with $\mathcal{E}$ replaced by $\mathcal{B}$.

\begin{figure}[h]
\includegraphics[width=0.48\textwidth]{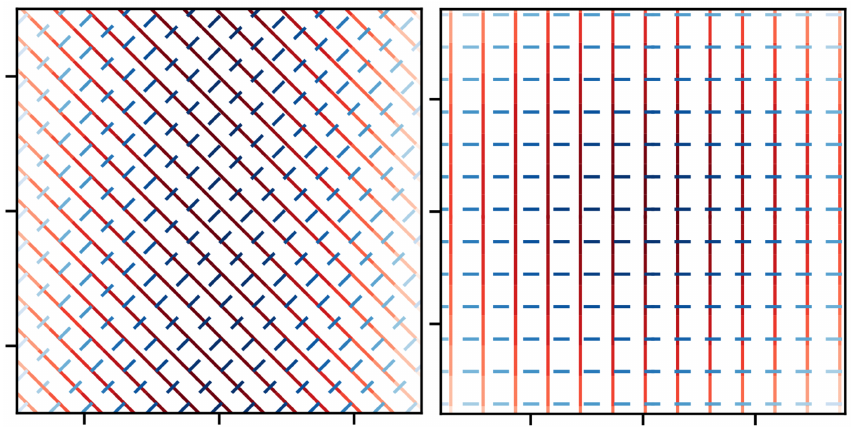}
\caption{\label{dit}
(color online).
%Tendexes in the case of \mbox{$\ell=1$} (left) and \mbox{$\ell=2$} (right) for small values of $r$.
%Tendexes  in the case of \mbox{$\ell=1$} (left) correspond to the eigenvectors ${\bf v}_1$ and ${\bf v}_2$ perpendicular to~${\bf v}_3$, see Eq.~\eqref{vvt11},
%which is almost aligned along $x$ axis in the range of coordinates examined - the step of the ticks is $0.03$.
%The above described orientation of the normal ${\bf v}_3$ means the section displayed corresponds to the $(y, z)$ plane.
%Values of the other parameters are \mbox{$t=x=0$} and \mbox{$k=\kappa=1$}.
%Similar figure also describes vortexes in the $(x, y)$ plane corresponding to the case shown on the right side. It shows how the vortex lines are rotated by~\mbox{$\pi/4$}. 
%Tendexes in the case of \mbox{$\ell=2$} (right) correspond to the eigenvectors perpendicular to a~normal almost aligned along $z$ axis in the range of coordinates examined, i.e., the $(x, y)$ plane is shown (the step of the ticks is $0.5$).
%Values of the other parameters are \mbox{$t=z=0$} and \mbox{$k=\kappa=1$}.
Left: Tendexes for \mbox{$\ell = 1$} in $(y,z)$ plane with \mbox{$t = x = 0$} and tick distance $0.03$. A similar plot represents vortexes for \mbox{$\ell = 2$} in $(x,y)$ plane with \mbox{$t = z = 0$}.
Right: Tendexes for \mbox{$\ell = 2$} in $(x,y)$ plane with \mbox{$t = z = 0$} and tick distance $0.5$.
Thus the vortex lines are rotated by~\mbox{$\pi/4$} with respect to the tendexes.
We put \mbox{$k = \kappa = 1$} in all cases.
Each panel is centered at the origin of the corresponding plane.
}
\end{figure}

The left panel of~Fig.~\ref{dit} shows tendexes given by Eq.~\eqref{vvt11}.
These tendex lines of ${\bf v}_1$ and ${\bf v}_2$ lie in a~plane described by a~normal ${\bf v}_3$; see Eq.~\eqref{vvt12}.
The normal is oriented along the $x$ axis for small values of~$r$ and for the chosen parameters.
The $z-$dependence, through $\phi$, in Eq.~\eqref{vvt12}, implies that the normal vector field -- approximately constant in the region displayed -- also rotates clockwise with increasing $z.$

\begin{figure}[h]
\includegraphics[width=0.48\textwidth]{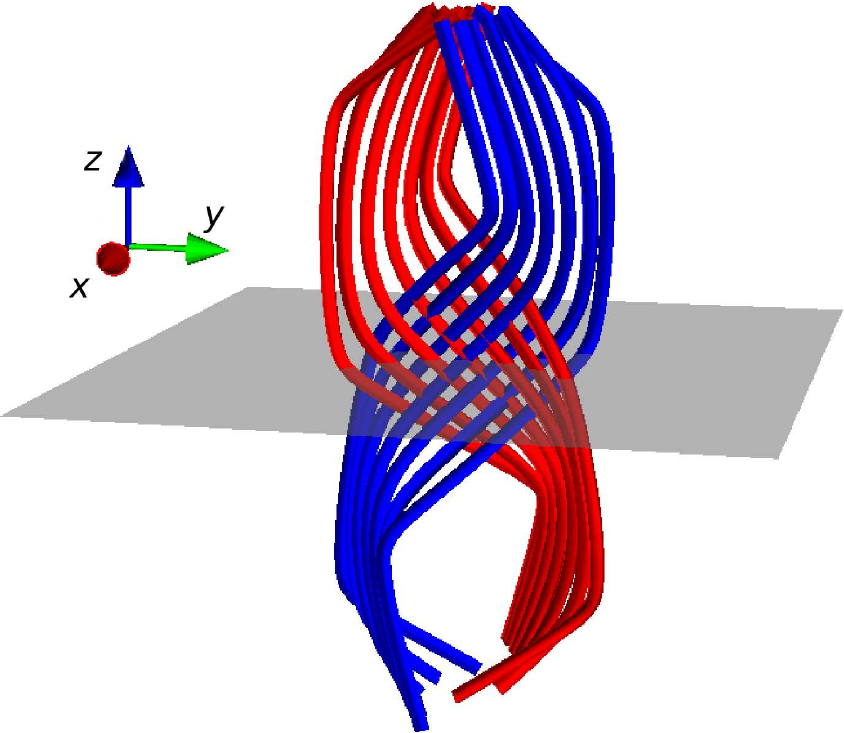}

\caption{\label{dit3D}
(color online).
Tendexes in the case of \mbox{$\ell=1$} for \mbox{$t = 0$} and \mbox{$k = \kappa = 1$}.
Tendex lines are color coded by the
sign of the eigenvalues, the most positive tendicity is colored blue (dark gray -- in the upper right and lower left) and the most negative tendicity is colored red (light gray -- in the upper left and lower right).
The range displayed along each axis is $[-3.5, 3.5]$.
}\end{figure}

More insight can be gained by a~pointwise exact solution, i.e. including terms of all orders in~$r$ in Eq.~\eqref{tendexes.bessel}.
Figure~\ref{dit3D} displays such an exact solution -- tendexes corresponding to the largest and to the lowest eigenvalues.
Their behavior is depicted in a~larger range than the one used in the left panel in Fig.~\ref{dit}.
One can clearly see that the pattern near the origin of the $(y, z)$ plane shown in Fig.~\ref{dit} (left) is present (close to the center of the shaded plane, in Fig.~\ref{dit3D}, but perpendicular to it) as expected.
The tendex lines leave the $(y, z)$ plane further away from the $x$-axis and they form a~helix-like pattern.

%Vortex lines (not displayed here) exhibit a~similar pattern as shown in~the left panel of~Fig.~\ref{dit} with red and blue lines interchanged.
%The plane of vortexes with the positive and negative vorticities is given
Vortex lines with the positive and negative vorticities (not displayed here) exhibit a~similar pattern as shown in the left panel of~Fig.~\ref{dit} with red and blue lines interchanged.
The plane of these vortexes (near the origin of the coordinate system) is given by a~normal aligned with y axis for parameters chosen as in Fig.~\ref{dit}.

%Similarly, the vortexes are given by
%$$\lambda_{1, 2} = \pm\sqrt{\mathcal{B}_{\hat{1}\hat{3}}^2 + \mathcal{B}_{\hat{2}\hat{3}}^2},\;
%\lambda_3 = 0$$
%and
%\begin{eqnarray}
%\sqrt{2}{\bf v}_{1, 2} &=& \frac{\mathcal{B}_{\hat{1}\hat{3}}}{\lambda_{1, 2}}\bfe{1} + \frac{\mathcal{B}_{\hat{2}\hat{3}}}{\lambda_{1, 2}}\bfe{2} + \bfe{3},\\
%{\bf v}_3 &=& \frac{-\mathcal{B}_{\hat{2}\hat{3}}\bfe{1}+\mathcal{B}_{\hat{1}\hat{3}}\bfe{2}}{\sqrt{\mathcal{B}_{\hat{1}\hat{3}}^2+\mathcal{B}_{\hat{2}\hat{3}}^2}}.
%\label{vvv0}
%\end{eqnarray}

\subsection{\label{sec:analysis.l2.r0}The case of \mbox{$\ell\geq 2$}, small values of $r$}
The range of the coordinate $r$ allows us to approximately solve for tendexes as follows
$$\lambda_{1, 2} = \pm\sqrt{\mathcal{E}^2_{\hat{1}\hat{1}}+\mathcal{E}^2_{\hat{1}\hat{2}}},\;
\lambda_3 = 0$$
and
\begin{equation}
{\bf v}_{1, 2} = \frac{(\lambda_{1, 2} + \mathcal{E}_{\hat{1}\hat{1}})\bfe{1} + \mathcal{E}_{\hat{1}\hat{2}}\bfe{2}}{\sqrt{2\lambda_{1, 2}(\lambda_{1, 2} + \mathcal{E}_{\hat{1}\hat{1}})}},\;
{\bf v}_3 = \bfe{3}.
\label{vvt2}
\end{equation}
Again, formulas for the vortexes are given by simply replacing $\mathcal{E}$ with $\mathcal{B}$ in the eigenvalue and eigenvector equations above.

%The case \mbox{$\ell=2$} is probably the most physically interesting.
A simple pattern of tendexes is displayed in~the right panel of~Fig.~\ref{dit}.
The pattern rotates clockwise with increasing~$z$ and counterclockwise with increasing~$t$ (which enters through the phase~$\phi$).

We shall illustrate this behavior using also a~standard approach to describe spacetime curvature effects -- geodesic deviation.
Consider a~circle of testing particles in the $(x, y)$ plane of a~small diameter with the center at the $z$ axis.
The tendex patterns indicate that this circle should be stretched along the red (solid) tendex lines and squeezed along the blue (dashed) tendex lines, we should get an ellipse that rotates as time flows.

Now, let us turn our attention to another well-established treatment of gravitational waves.
The equation of geodesic deviation [Eqs.~(35.14) and (35.15) of MTW~\cite{mtw}] in the proper reference frame~\cite{honza} describes the oscillations of locations of test particles which are induced by the wave as measured by an observer in the spatial origin of the proper reference frame.
The equation is often solved in a~convenient transverse-traceless ($TT$) gauge where \mbox{$h^{TT}_{jk}=2\omega^{-2}R_{j0k0}$} (Box $35.1$ in MTW~\cite{mtw}).
We do not present the explicit and rather lengthy expressions of the gravitational Bessel wave in the transverse-traceless gauge, because they are the same (up to the constant~$2\omega^{-2}$) as the expressions for the tidal field in~Eq.~\eqref{tendexes.bessel} as follows from~Eq.~\eqref{tidal.general}.

\begin{figure}[h]
\includegraphics[width=0.48\textwidth]{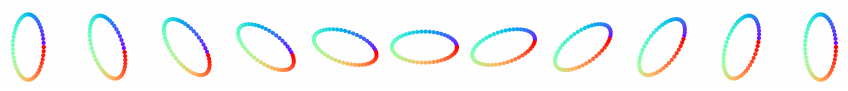}
\caption{\label{rotpol}
(color online).
Time sequence depicting the transverse effect of the gravitational Bessel wave with \mbox{$\ell=2$} in the $(x, y)$ plane at equidistant times over one period.
}
\end{figure}

If we consider the limit \mbox{$r \rightarrow 0$}, we get nonzero components of the $TT$-gauge metric only for~\mbox{$\ell\leq 2$}.
Figure~\ref{rotpol} shows the time evolution of the $(x, y)$ plane oscillations (\mbox{$z=0$}), perpendicular to the direction of the propagation of the vortex wave for the case~\mbox{$\ell=2$}.
The consecutive time snapshots reveal a~rotating ellipse.
This agrees with the rotating polarization in the right-handed direction as predicted by the tendexes, i.e., counterclockwise for a~gravitational wave propagating toward the reader.
\begin{figure}[h]
\includegraphics[width=0.48\textwidth]{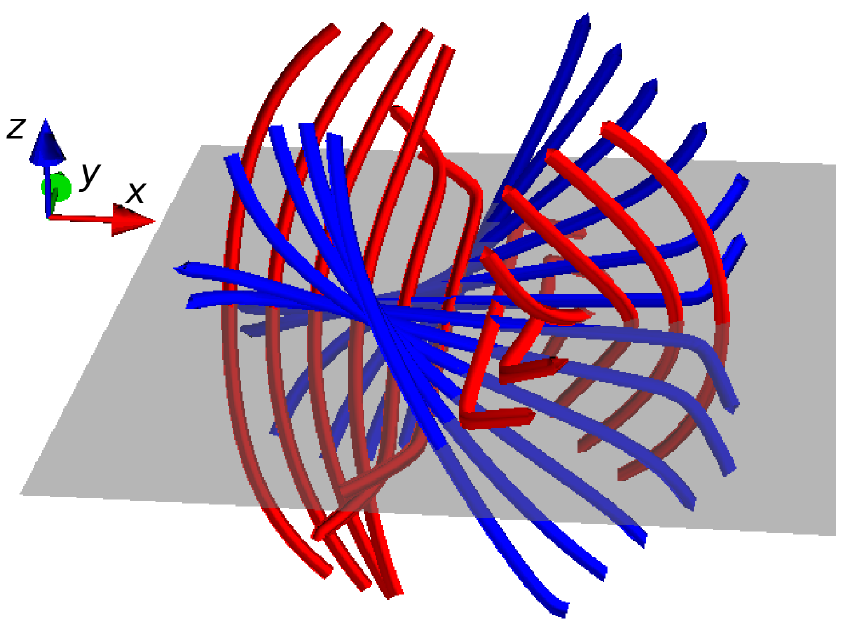}
\caption{\label{quad3D}
(color online).
Tendexes in the case of \mbox{$\ell=2$} with \mbox{$t = 0$} and \mbox{$k = \kappa = 1$}.
Tendex lines are colored by the sign of the eigenvalues, blue (dark gray) for the most positive and red (light gray) for the most negative.
The range of each of the three coordinates is $[-3.5, 3.5]$.
}\end{figure}

Figure~\ref{quad3D} displays the tendexes corresponding to the largest and the lowest tendicities in order to describe their behavior beyond the approximation used in the right panel in Fig.~\ref{dit}.
One can see that the corresponding part of Fig.~\ref{dit} -- the perpendicular set of lines oriented along the $x$, respectively $y$, axis -- is reproduced near the origin (center of the shaded plane) as expected.
The tendexes leave the $(x, y)$ plane further away from the $z$ axis and a~spiraling pattern %along the $z$-axis
is revealed.

\begin{figure}[h]
\includegraphics[width=0.48\textwidth]{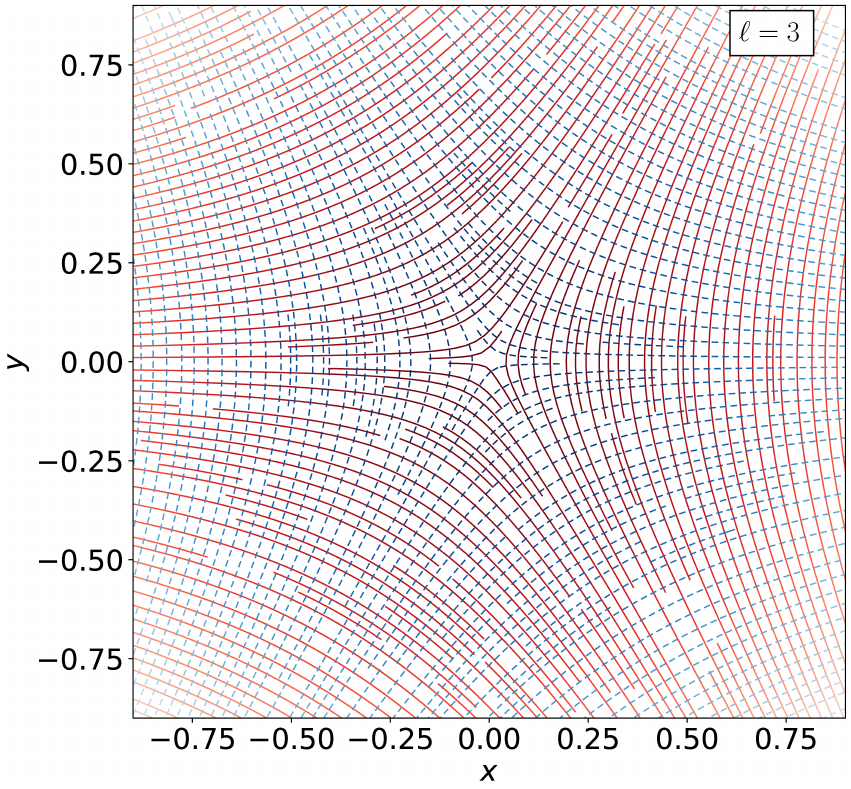}
\caption{\label{trit}
(color online).
Tendex lines in the $(x, y)$ plane with the same parameters as in the right panel of~Fig.~\ref{dit} but in the case of \mbox{$\ell=3$}.}
\end{figure}

In Fig.~\ref{trit} we present the structure of tendex lines in the $(x, y)$ plane for the case~\mbox{$\ell=3$}.
The diagram demonstrates the appearance of a~singular point in the center.
Singular points are also present for higher values of $\ell.$
Such a~point is called the triradius if~\mbox{$\ell=3$}.
The appearance of these singular points is related to the vanishing of components of the $TT$-gauge metric (and thus also the tidal field) in the limit \mbox{$r \rightarrow 0$} for~\mbox{$\ell\geq 3$}.

Directions of vortex lines corresponding to the case~\mbox{$\ell=2$} are indicated in the left panel of~Fig.~\ref{dit}. 
It demonstrates that the vortexes are rotated by~\mbox{$\pi/4$} with respect to the tendexes.
The magnitude of vorticities has the same decaying behavior as the magnitude of tendicities.
The vortex lines in the case \mbox{$\ell=3$} are rotated by~\mbox{$\pi/6$} with respect to the tendex lines
in the close vicinity of the center.

Let us note that \mbox{$\ell\geq 2$} tendexes of positive and negative eigenvalue exhibit the same pattern which is only rotated by \mbox{$\pi/\ell$} close enough to the $z$ axis.
The same applies to vortexes.

\subsection{\label{sec:analysis.l.r.large}Asymptotical behavior for large values of $r$}

The Newman-Penrose formalism is an approach widely used to analyze the asymptotical behavior of a~gravitational wave at null infinity. 
Let us recall the basic ingredients.
We use the following complex null tetrad
\begin{eqnarray*}
{\bf l} = \frac{1}{\sqrt{2}}\left(\bfe{0} + \bfe{1}\right)
&,\quad &
{\bf n} = \frac{1}{\sqrt{2}}\left(\bfe{0} - \bfe{1}\right)
,\\
{\bf m} = \frac{1}{\sqrt{2}}\left(\bfe{2} + i\bfe{3}\right)
&,\quad &
{\bf m}^{\ast} = \frac{1}{\sqrt{2}}\left(\bfe{2} - i\bfe{3}\right)
.\\
\end{eqnarray*}
We inserted the tetrad from Eq.~\eqref{ourtetrad} into the above formulas and we computed the complex
Weyl scalars defined as follows
\begin{eqnarray*}
\Psi_0 &=& C_{\mu\nu\rho\sigma}l^{\mu}m^{\nu}l^{\rho}m^{\sigma}
,\ 
\Psi_1 = C_{\mu\nu\rho\sigma}l^{\mu}n^{\nu}l^{\rho}m^{\sigma}
,\\
\Psi_2 &=& C_{\mu\nu\rho\sigma}l^{\mu}m^{\nu}m^{\ast\rho}n^{\sigma}
,\ 
\Psi_3 = C_{\mu\nu\rho\sigma}l^{\mu}n^{\nu}m^{\ast\rho}n^{\sigma}
,\\
\Psi_4 &=& C_{\mu\nu\rho\sigma}n^{\mu}m^{\ast\nu}n^{\rho}m^{\ast\sigma}
.\\
\end{eqnarray*}
The authors of~\cite{nichols} (see their Appendix) have proved a~relation
among the tidal field $\mathcal{E}$, the frame-drag field $\mathcal{B}$ and the Weyl scalars
\begin{equation}
\mathcal{E}_{\hat{a}\hat{b}}+i\mathcal{B}_{\hat{a}\hat{b}} = 
\begin{bmatrix}
2\Psi_2 &
-\left(\Psi_1 - \Psi_3\right)&
i\left(\Psi_1 + \Psi_3\right)
\\[\matrixlinesep]
\ast &
\frac{\Psi_0 + \Psi_4}{2} - \Psi_2 &
-\frac{ i}{2}\left(\Psi_0 - \Psi_4\right)
\\[\matrixlinesep]
\ast &
\ast &
-\frac{\Psi_0 + \Psi_4}{2} - \Psi_2
\\[\matrixlinesep]
\end{bmatrix}
.\label{psi.bessel}
\end{equation}

The peeling theorem~\cite{newmanpenrose} leads to the conclusion that asymptotically only~$\Psi_4$ contributes to~$\mathcal{E}$ and~$\mathcal{B}$ in an asymptotically flat spacetime.
The gravitational Bessel wave does not decay in the~$z$ direction; hence our spacetime under study -- the Minkowski spacetime perturbed by the wave -- is not asymptotically flat in the strict sense required by the peeling theorem.
Also the radial falloff~$\sim r^{-\frac{1}{2}}$ is slower than for an asymptotically flat spacetime.
Nevertheless, the general formula~\eqref{psi.bessel} does still apply.
%for our tetrad~\eqref{ourtetrad}.
We obtained that in leading order (in~$r$) only the even-numbered Weyl scalars are real and the 
odd-numbered Weyl scalars are imaginary.
Then Eq.~\eqref{psi.bessel} implies that the tidal field~$\mathcal{E}$ is given by the entries on the main and the minor diagonal and the frame-drag field~$\mathcal{B}$ is given by the complementary entries.
The leading order behavior in Eqs.~\eqref{tendexes.bessel} and~\eqref{vortexes.bessel} confirms this result.

\begin{figure}[h]
\includegraphics[width=0.48\textwidth]{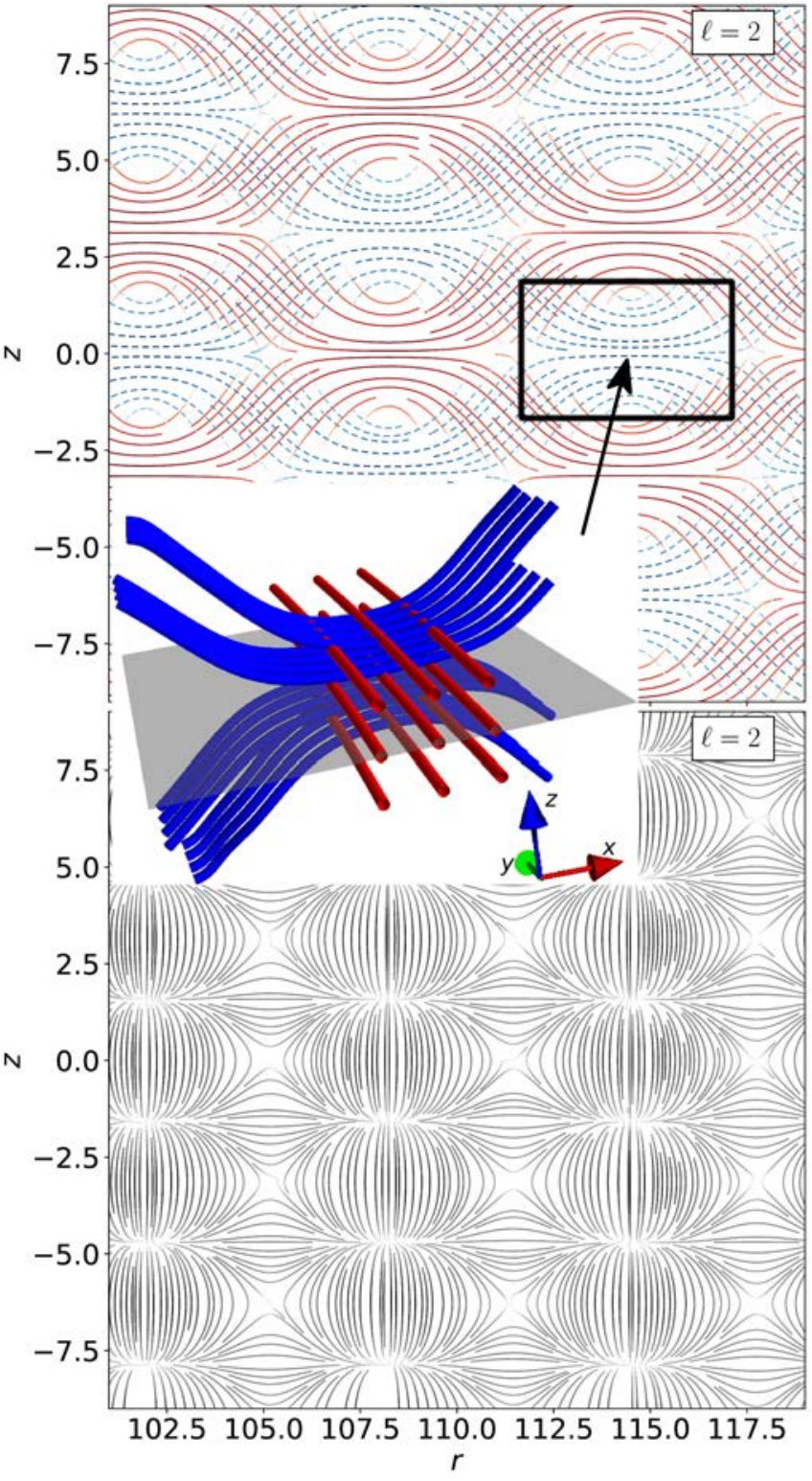}
\caption{\label{quat.a}
(color online).
Tendexes and vortexes in the case of \mbox{$\ell=2$} and large values of~$r$ corresponding to the eigenvector(s) in the $(r, z)$ plane ($t=0$, \mbox{$\varphi=0$} and \mbox{$k=2\kappa=1$}).
\\
Top and inset:
The top panel displays positive-tendicity lines as dashed (blue) and negative-tendicity ones as solid (red) lines.
The 3D inset %in the middle
contains the minimal tendicity lines -- in red (light gray, almost aligned with the $y$ axis) -- and
the maximal tendicity lines -- in blue (dark gray).% in a~zoomed area marked by a~black rectangle.
\\
Bottom:
Vortex lines of identically zero vorticity.
The remaining two vortex lines (not displayed) cross the lines shown, but are not tangent to the plane, and they have vorticities that differ only in sign.
Shading of the curves presented in this plot is determined by the absolute value of the vorticity of the other two vortexes.
}
\end{figure}

This implies that the limit for large values of~$r$ must have formally the same solution of the eigenvalues and eigenvectors as the case of \mbox{$\ell=0$}.
Moreover, the formal solution is applicable to all values of $\ell$; this holds true for both tendexes and vortexes.
Figure~\ref{quat.a} (top) displays the tendex lines in~the $(r, z)$ plane for \mbox{$\ell=2$} and $k\neq\kappa$.
A~periodic tiling, similar to that in~Fig.~\ref{monot} (with \mbox{$\ell=0$}), appears again.
The choice \mbox{$k\neq\kappa$} produces wiggly tendexes.
This contrasts the tendex lines that are resembling straight lines almost everywhere in the case of \mbox{$k=\kappa$} (not displayed in a~separate figure but already forming for larger $r$ in~Fig.~\ref{monot}).
We display a~3D inset of tendexes instead of periodic tiling in the bottom part of Fig.~\ref{monot}.
The most positive/negative tendicity lines are shown.
We see that the red curves, no longer visible in the $(r, z)$ plane in the vicinity of the point marked by the arrow, are aligned with the direction tangent to the coordinate $\varphi$.
The reason why this occurs is the same as the one already mentioned in Sec.~\ref{sec:analysis.l0.r}.
In Fig.~\ref{quat.a} (bottom) the corresponding vortex lines are also shown.
The same comments as in the case of \mbox{$\ell=0$} apply here.
Moreover, the features displayed in Fig.~\ref{quat.a} are qualitatively identical for higher values of $\ell$.
Of course, the dependence on $\varphi$ will be different for different values of~$\ell$ but it is not very significant in Fig.~\ref{quat.a}.

Let us comment on the direction of propagation of the wave.
Consider the enlarged region in Fig.~\ref{quat.a} at the tip of the arrow.
The zero vorticity line (bottom panel) is vertical there.
The $(r, z)$ section displayed is a~plane of reflection symmetry for the remaining mutually perpendicular vortexes, thus these vortex lines intersect the plane of the figure with inclination angle equal to \mbox{$\pi/4$} and they are orthogonal to the vertical curve shown.
As a~result they must form a~``cross~$\times$" pattern in the $(x, y)$ plane.
The positive tendicity lines are aligned along the $x$ direction %(represented by the $r$ axis in the case of \mbox{$\varphi=0$})
while the negative tendicity lines are aligned along the $y$ direction, i.e., a~``plus~$+$" pattern is formed.
This is clearly seen from both the top panel and the inset.
In summary, the vortex and tendex lines are $\times$ shaped, coplanar and relatively rotated by \mbox{$\pi/4$}.
This implies that the wave is locally propagating along the perpendicular direction, the $z$ axis.
Thus the situation depicted in Fig.~\ref{dit} close to \mbox{$r=0$} is locally reproduced in the large $r$ regions as well.
The behavior of the wave is much more complex in a~general position.

It is known that wavefronts of Durnin's photon beams have a~shape of a~generalized cone; see e.g.~\cite{ssanchez}.
In our case, the tiling shifts in the $z$ direction as we change the angle~$\varphi$.
Thus we can suspect that a~sort of generalized cone (wiggly for $k\neq\kappa$) is present also in the case of a~gravitational Bessel wave.
A deeper analysis reflecting an impact of Poincar\'{e}-Hopf theorem as in~\cite{zimmerman} with a~sphere replaced probably by a~conelike surface would be necessary to reach a~more definite conclusion.

\subsection{\label{sec:analysis.0.limits}The case of \mbox{$k=0$}}

In this subsection, we will discuss the special case of \mbox{$k=0$}, which for \mbox{$\ell=0$} corresponds to cylindrical gravitational waves extensively studied in the literature.
This knowledge can give us some hints of what to expect in our general case, e.g., a~type of possible sources.
We shall recall some related properties of cylindrical gravitational waves and then we will turn our attention to the specific subcase of our solution.

Cylindrically symmetric waves are locally characterized by two commuting, spatial, Killing vector fields with an integrable orthogonal space.
One can find coordinates $(\varphi, z)$ such that the Killing
vectors are \mbox{$\partial/\partial\varphi$} and \mbox{$\partial/\partial z$}.
Furthermore, there exists an axis with respect to the coordinate $\varphi$ (with $0$ and $2\pi$ identified)~\cite{hayward}. %see Hayward (CQG)

Cylindrically symmetric waves have played a~significant role in the solution of several problems.
We will mention some of the problems in the following.
There is a~well-known exact solution of Einstein equations -- the already mentioned Einstein-Rosen waves, which were utilized in discussions concerning the reality of gravitational radiation~\cite{weberwheeler}.
Results about the underlying topology of a~specific class of cylindrical gravitational waves are available as well~\cite{gowdyedmonds}, e.g., spacelike hypersurfaces of these gravitational waves can be homeomorphic for example to a~tree-torus.
This means that these waves admit compact sources.
Moreover, waves of this kind belong to the class of boost-symmetric spacetimes~\cite{bicak}.

The importance of general cylindrically symmetric waves naturally leads to the search for the corresponding matter sources, even if they are often idealizations.
These include a~pulsating mass cylinder~\cite{marder} and cosmic strings~\cite{anderson}.
Cosmic strings were proposed by Kibble as one-dimensional topological defects during a~symmetry breaking phase transition early after the big bang~\cite{kibble}.
An analog of the black-hole thermodynamic laws can be formulated also for cylindrically symmetric cosmic strings.
They were studied in~\cite{hayward}, e.g., their modified Thorne energy, and the energy flux of the gravitational waves, among other invariantly defined quantities.
Cosmic strings can generate gravitational waves in broad bands from low to high frequencies~\cite{wen}.

The \mbox{$k=0$} case is simple enough to be solved without any approximation in $r$ in order to analyze the vortexes and tendexes.
Moreover, the components in Eqs.~\eqref{tendexes.bessel} and~\eqref{vortexes.bessel} are independent of $z.$

\begin{figure}[h]
\includegraphics[width=0.48\textwidth]{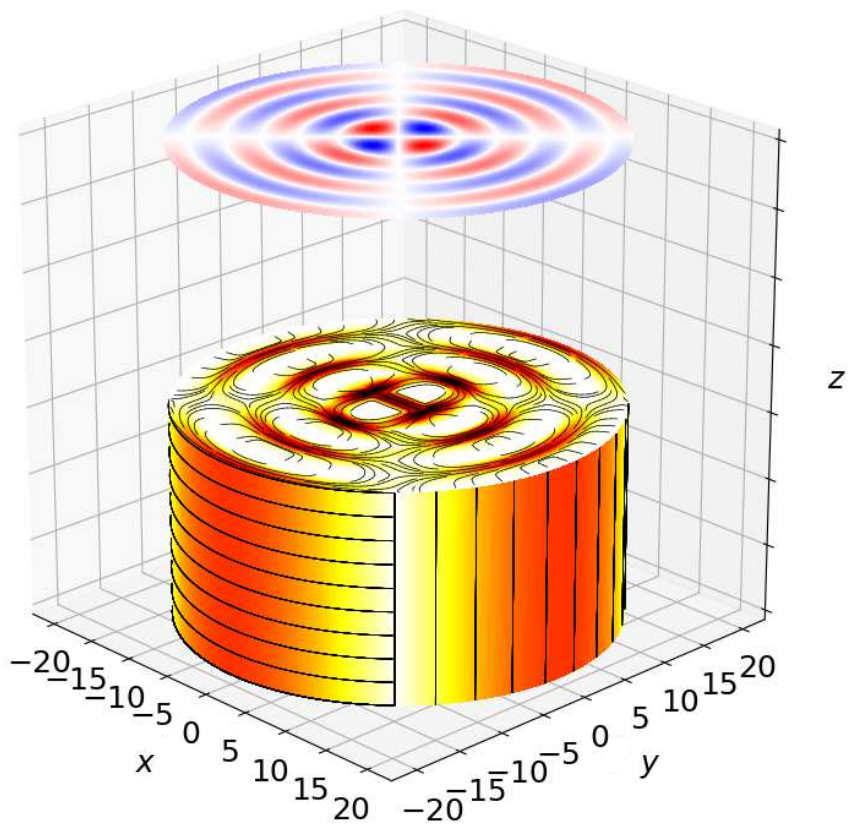}
\caption{\label{quat3d.zero.k.a}
(color online).
Tendex lines in the case of \mbox{$\ell=2$}, \mbox{$k=0$}.
The figure consists of the following parts:
The top part displays tendicity corresponding to the vector field \mbox{$\partial/\partial z.$}
The red (lighter gray) areas have negative tendicity and the blue (dark gray) areas have positive tendicity.
The color intensity represents the magnitude of the eigenvalue.
The cylinder displays positive tendex lines also shown in~Fig.~\ref{quat.zero.k.a}.
This time, the tendicity is marked using a~background color map instead of line color; purple
(darker) regions correspond to a~large eigenvalue, and yellow (lighter) regions depict a~value closer to zero.
The side of the cylinder displays large $r$ limit of the tendexes evaluated at the radius of the cylinder.
}
\end{figure}

The tidal field in Eq.~\eqref{tendexes.bessel} is a~block-diagonal matrix composed of two consecutive matrices, a 2$\times$2 matrix and a 1$\times$1 matrix, along the main diagonal.
This means that the coordinate vector field \mbox{$\partial/\partial z$} is always an eigenvector with a~corresponding eigenvalue given by $\mathcal{E}_{\hat{3}\hat{3}}.$
The formula for the eigenvalue shows that its radial dependence is given by the oscillating Bessel function.
It follows that the plot will consist of concentric regions where the tendicity changes its sign when we move from one region to its neighbor along the radial direction.
In other words, the color of the displayed tendicity will oscillate between red and blue with changing $r.$
The $\varphi$ dependence, through $\cos\phi$, will also lead to azimuthal oscillations provided \mbox{$\ell\neq 0$}.
In the case of \mbox{$\ell=0$}, only concentric annuli, each of a single color, appear.
An example in the case of \mbox{$\ell=2$} is displayed in the top part of Fig.~\ref{quat3d.zero.k.a}.

\begin{figure}[h]
\includegraphics[width=0.48\textwidth]{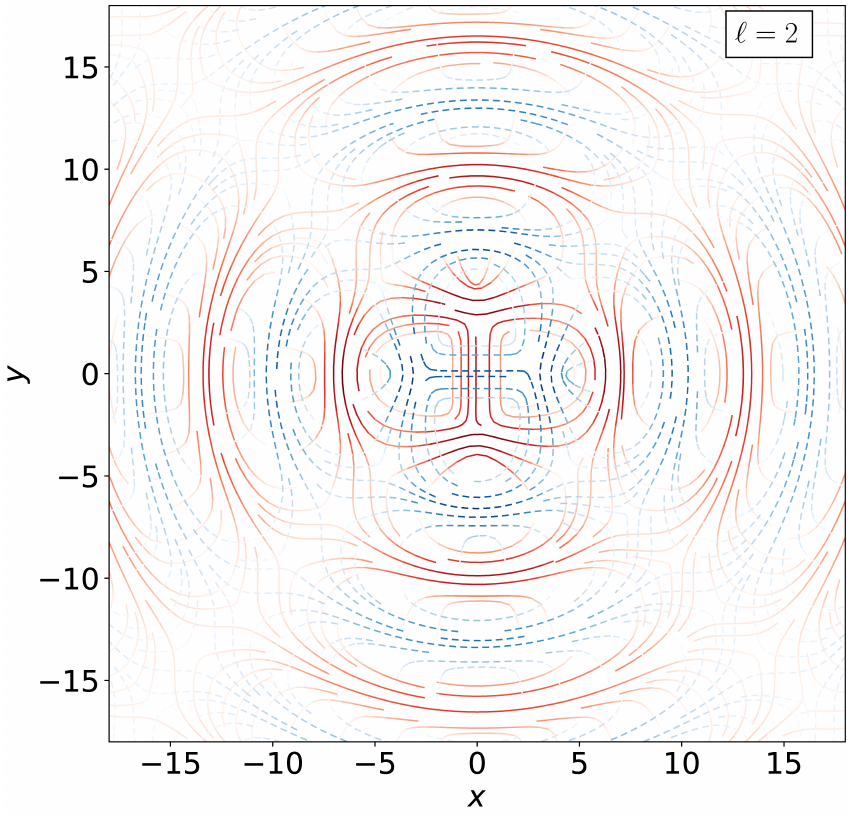}
\caption{\label{quat.zero.k.a}
(color online).
Tendexes in the case of \mbox{$\ell=2$}, \mbox{$k=0$} (exact in $r$) corresponding to the eigenvectors in the $(x, y)$ plane in the range of coordinates examined.
Values of the other parameters are \mbox{$t=0$} and \mbox{$\kappa=1$}.}
\end{figure}

\begin{figure}[h]
\includegraphics[width=0.48\textwidth]{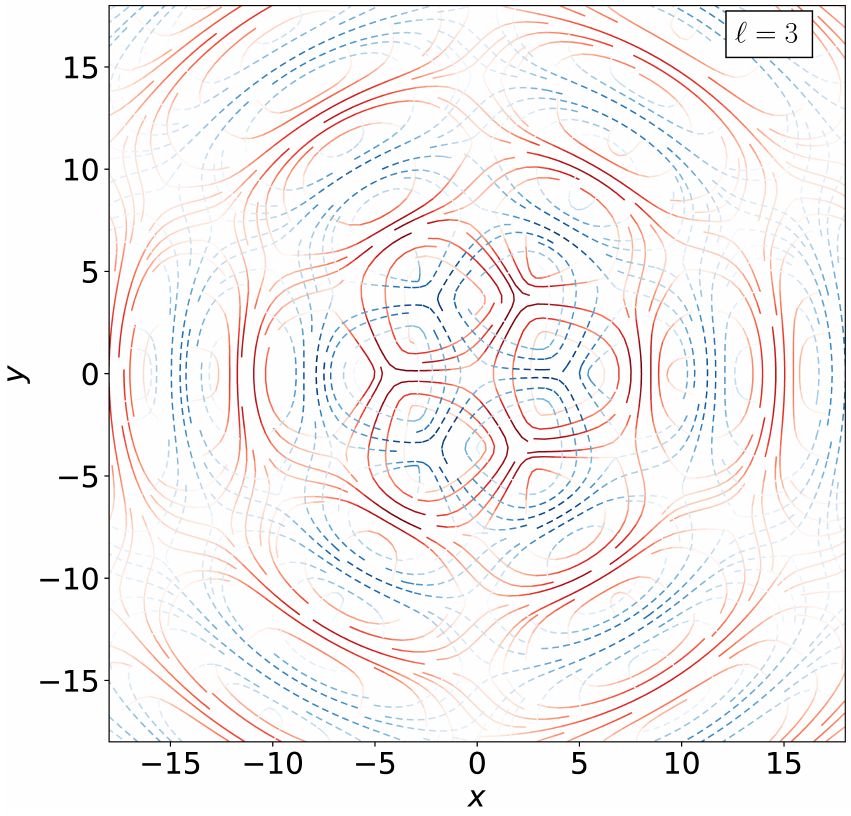}
\caption{\label{octu.zero.k.a}
(color online).
Tendexes in the $(x, y)$ plane for the same parameters as in~Fig.~\ref{quat.zero.k.a} but in the case of \mbox{$\ell=3$}.
}
\end{figure}

The behavior of the two remaining tendexes is more complex as far as $\ell$ is concerned.
Even so, periodicity in the azimuthal direction is present here as well, compare the case of \mbox{$\ell=2$} and \mbox{$\ell=3$} in Figs.~\ref{quat.zero.k.a} and~\ref{octu.zero.k.a}.

The case of \mbox{$\ell=0$} represents a~cylindrically symmetric wave as already mentioned.
The vanishing of $\ell$ ensures that the $\varphi$ dependence of components in Eq.~\eqref{tendexes.bessel} drops out.
The tidal field in Eq.~\eqref{tendexes.bessel} becomes diagonal which implies that the eigenvectors are the coordinate vector fields.
Plots of each of the two eigenvectors in the $(x, y)$ plane will be composed from concentric annuli regions of oscillating sign of tendicity.
Of course, the positions where the sign changes are specific to each of them due to different radial dependence of the eigenvalues.
Nevertheless, we can make a~general comment on the asymptotic behavior because the radial tendicity fades out faster than the other two eigenvalues.
This means that the tendicities corresponding to \mbox{$\partial/\partial\phi$} and \mbox{$\partial/\partial z$} are of the same magnitude but opposite sign for sufficiently large values of $r$.
The cylindrical wave becomes transversal and it turns into a~plane wave locally, with the propagation direction \mbox{$\partial/\partial r$}, far away from the $z$ axis.

A similar reasoning applies to the case of \mbox{$\ell>0$}: the radial tendicity fades away faster again.
The situation is depicted in Figs.~\ref{quat3d.zero.k.a} and~\ref{quat.zero.k.a} in the case of \mbox{$\ell=2$}.
The side of the cylinder of Fig.~\ref{quat3d.zero.k.a} describes the asymptotical behavior of positive tendex lines.
Because the remaining tendicities of the two eigenvectors \mbox{$\partial/\partial\phi$} and \mbox{$\partial/\partial z$} sum up to zero and the tendicities oscillate, there are regions where positive tendicity is associated with \mbox{$\partial/\partial\phi$} and regions where it is associated with \mbox{$\partial/\partial z$}.
Figure~\ref{quat3d.zero.k.a} displays this fact by the presence of alternating regions in the cylinder side -- regions with horizontal tendex lines and regions with vertical tendex lines.
The boundaries of these regions coincide with the ``rays" of vanishing tendicity (white regions) corresponding to the eigenvector \mbox{$\partial/\partial z$} depicted in the topmost part of Fig.~\ref{quat3d.zero.k.a}.
We also plot the magnitude of the tendicity on the cylinder, we use a~color map, as
in~\cite{zimmerman}, in which purple (darker) areas correspond to large eigenvalues and yellow (lighter) regions are closer to zero.

The top base of the cylinder in Fig.~\ref{quat3d.zero.k.a} contains tendexes with the positive tendicity for finite~$r$.
The side is joined to the base as if it were at infinity.
The topmost part of the figure shows that the positive tendicity is transferred from the white regions of the % $(x, y)$
horizontal plane to the vertical direction with positive tendicity indicated by a~blue color.
The base is presented in more detail in~Fig.~\ref{quat.zero.k.a} including tendexes with negative tendicity.
One can see that the plot of tendexes bears a~striking similarity to tendexes of an oscillating mass quadrupole, see Fig.~15 of~\cite{nichols}, except for the central region which is smoothed out in our case.
The radial tendicity is strong near the center \mbox{$r=0$} but diminishes further away from the $z$ axis.
The near $z$-axis zone contains perpendicular tendex lines of opposite sign, effectively reproducing the right panel of~Fig.~\ref{dit}.
As one moves away from the center, the tendex lines form distorted closed loops that stay in a~single quadrant delimited by lines \mbox{$y=\pm x$} of the displayed plane.
Still, there are some tendexes crossing quadrants near triradius points.
%Still, there are some tendexes that pass around these closed loops in the following fashion:
%they approach the outer crest of a~loop (larger value of $r$), pass near a~triradius point and they approach the inner crest of a~neighboring loop.
%Such lines pass through all the quadrants.

\begin{figure}[h]
\includegraphics[width=0.48\textwidth]{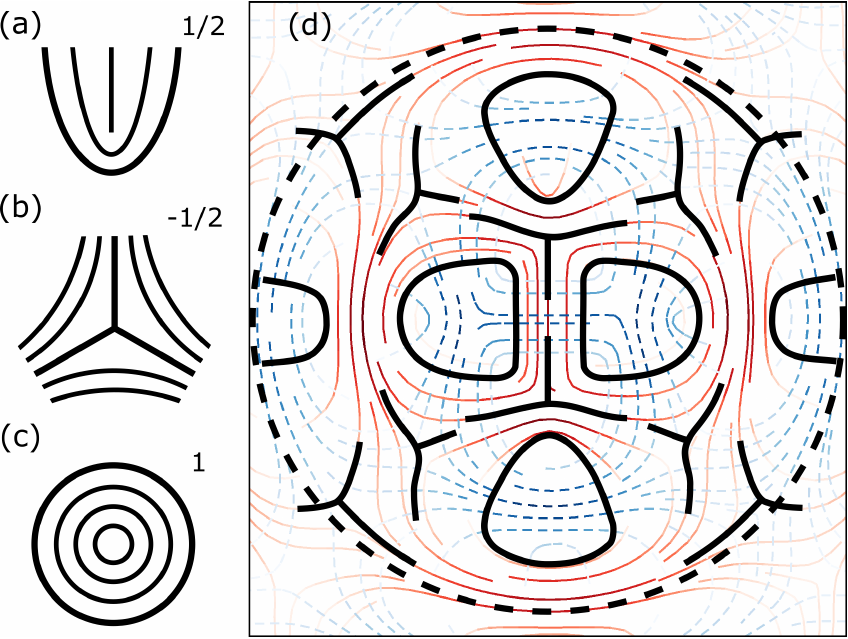}
\caption{\label{euler.quat.zero.k.a}
(color online).
The left column displays sketches of the three basic singularities together with values of the corresponding indices: (a) open loop \mbox{$i=1/2$}, (b) triradius \mbox{$i=-1/2$} and (c) closed loop \mbox{$i=1$}.
The right panel (d) is an~enlargement of the central area in Fig.~\ref{quat.zero.k.a} with these singularities explicitly marked in the case of negative tendicity lines (red, solid) and within a~dashed line circle.}
\end{figure}

The cylinder in~Fig.~\ref{quat3d.zero.k.a} (including the circular end caps) is homeomorphic to a~sphere; therefore it must have the same Euler characteristic \mbox{$\chi=2$}.
White regions in~Fig.~\ref{quat3d.zero.k.a} can be shrunk, tendex lines can be radially extended and the indices $i$ of all isolated singularities can be calculated.
The Poincar\'{e}-Hopf theorem requires that the contributions coming from indices $i$ corresponding to open loops \mbox{$i=1/2$} [Fig.~\ref{euler.quat.zero.k.a}(a)], triradii points \mbox{$i=-1/2$} [Fig.~\ref{euler.quat.zero.k.a}(b)], and closed loops \mbox{$i=1$}, [Fig.~\ref{euler.quat.zero.k.a}(c)] (see~\cite{zimmerman,crowley}), do sum up to \mbox{$\chi=2$}.
Let us note that this result must be independent of the sign of the tendicities of the tendex lines.
%Because the Fig.~\ref{quat.zero.k.a} explicitly shows that these two patterns are relatively rotated, it suffices to perform the calculations for one of them only.
By counting the individual contributions of the singular sets lying within a~circle of a~given radius in each base of the cylinder, one can show that they always sum up to zero.
In order to illustrate this fact, Fig.~\ref{euler.quat.zero.k.a}(d) exhibits a~finite radius circle with the singularities explicitly marked in the case of negative tendicity lines for better visual presentation.
One can verify that their indices do sum up to zero.
This holds true even if the radius tends to infinity.
Now, only the singular sets in the side of the cylinder need to be taken care of.
They can be viewed as a~continuation of the open loops on the two bases.
It means that we can count each of these four singular lines as a~loop with the index \mbox{$i=1/2$}.
It follows that the total sum of the indices is equal to the Euler characteristic \mbox{$\chi=2$}.
Similarly, we can count indices in the case \mbox{$\ell=3$} with the tendexes in the $(x,y)$ plane shown in~Fig.~\ref{octu.zero.k.a}.
This time each base would contribute with the index~\mbox{$i=-1/2$} and each of the six singular lines on the side of the cylinder with the index~\mbox{$i=1/2$}, summing up to $2$ again.

%There are four of them, which means the total contribution is \mbox{$4\times1/2 = 2$}.

Now, let us briefly turn our attention to vortexes.
The solution is formally the same as for tendexes/vortexes in the case of \mbox{$\ell=1$} and small values of $r$; see Eqs.~\eqref{vvt11} and~\eqref{vvt12}.
When comparing the vortexes and tendexes, one can easily verify that the asymptotical behavior for large $r$, irrespective of the value of $\ell$, is characterized by the fact that the vortex and tendex lines are locally rotated by \mbox{$\pi/4$}.

\section{\label{sec:conclusions}Conclusions}
Motivated by the observation of gravitational waves and the usefulness of Bessel vortex beams in other fields of science,
we have found a~multidimensional Bessel beam inspired solution to linearized Einstein gravity.
The multidimensional solution, presented in this paper for the first time, fits a~broader research program exploring and testing alternative general relativity models.
The four-dimensional reduction is a~generalization of a~cylindrical gravitational wave.%~\cite{mtw}.

We were inspired by electron and photon Bessel vortex waves in looking for a~gravitational analog.
We employed the similitude between electromagnetism and gravitation also in utilization of the general relativity analogs of electric and magnetic field lines -- tendexes and vortexes.
We have applied the tendexes and vortexes to visualize the curvature of the four-dimensional solution.
These visualizations clarify that the waves propagation is not purely radial, axial, azimuthal or standing, in a~general case, but it includes a~rich interplay between all of these types of wave behavior.
We have found and discussed some similarities with the weak gravity phenomena examined in the series of papers pioneering these powerful visualization tools.
These similarities can serve as an indication to the experts in numerical relativity to pinpoint the potential sources of the gravitational Bessel wave.
Possible sources could be oscillating or interacting cosmic strings as mentioned in the discussion in Sec.~\ref{sec:analysis.0.limits}.
Also, the optical and/or acoustic resemblance suggests that such a~wave could be produced via scattering of a~gravitational plane wave on an appropriate analog of scattering centers mentioned in the Introduction.
Whatever the system producing the gravitational Bessel waves and its dynamics may be,
we are confident that the presented visualizations will support physical intuition in
analyzing further these waves and their sources.

We analyzed the solution for different choices of parameters and also its asymptotical
behavior. We discussed properties of the Weyl scalars for an appropriate tetrad and
for the cylindrical-like case we explored singularities of its tendex line patterns.
These singularities have to develop for such compact surfaces and they show a~fingerprint
of the beaming characteristics of the spacetime currently studied.

We hope that our paper and its findings will both contribute to and provide inspiration for the rapidly expanding research of gravitational waves.

\begin{acknowledgments}
A. P., M. Z. and T. \v{R}. acknowledge funding from the Technology Agency of the Czech Republic (Centre of Electron and Photonic Optics, Grant No: TN0100000).
O. B. acknowledges support from the Czech Science Foundation (project GA18-27546S) and MEYS %(Ministry of Education, Youth and Sports)
CZ.02.1.01/0.0/0.0/16\_026/0008460.
The research infrastructure was supported by MEYS
of the Czech Republic, the Czech Academy of Sciences,
and the European Commission (Grants No. LO1212,
No. RVO:68081731, and No. CZ.1.05/2.1.00/01.0017).
\end{acknowledgments}

\appendix
\section{\label{sec:general.D.solution}The general dimension linearized solution}
\subsection{\label{sec:general.D.ansatz}Conventions and the ansatz}
The background metric is expressed in the Minkowski coordinate system as follows
$$\left.\eta_{\mu\nu}\d x^{\mu}\d x^{\nu}\right|_{\rm Mink.} = - \d t^2 + \d l^2,\; \d l^2 = \delta_{uv}\d x^u\d x^v.$$
In the cylindrical-like one, the spatial part of the metric becomes
\begin{equation}
\d l^2 = \underbrace{\left[\d r^2 + r^2\d\Omega^2_{B-1}\right]}_{{\rm indices}\; a, b, \ldots} + \underbrace{\delta_{ij}\d x^i\d x^j}_{{\rm indices}\; i, j, k, \ldots}
,\label{space.split}
\end{equation}
where $\d\Omega^2_{B-1}$ is a~volume element of a~unit sphere of dimension $B-1.$
The angular coordinates range as follows
$$\vartheta^a\in[0, \pi],\ \text{for}\ a\in[2, B-1],\ \varphi\equiv\vartheta^B\in[0,2\pi)$$
The multidimensional sphere has a~nontrivial topology, so at least two coordinate charts
(e.g., using a~stereographic projection) are necessary to cover the entire manifold.
Since we leave these global questions aside, we will work only with one coordinate patch - with
hyperspherical coordinates, so we can parametrize the hypersphere of radius $r$ in $\mathbb{R}^B$ using Cartesian coordinates by
\begin{equation}\label{hypersphereparam}
\begin{split}
x^1 &= r\sin\vartheta^2\sin\vartheta^3\ldots\ldots\ldots\cos\vartheta^B,\\
x^2 &= r\sin\vartheta^2\sin\vartheta^3\ldots\ldots\ldots\sin\vartheta^B,\\
x^3 &= r\sin\vartheta^2\sin\vartheta^3\ldots\cos\vartheta^{B-1},\\
&\vdots\\
x^{B-1} &= r\sin\vartheta^2\cos\vartheta^3,\\
x^B &= r\cos\vartheta^2.
\end{split}
\end{equation}
The split of the spatial part in~\eqref{space.split} is of the same type as in~\cite{cermak.zouhar.2014,cermak.zouhar.2012}.

The multidimensional ansatz for the perturbation field is
\begin{equation}
h_{\mu\nu} = \Re\{ A_{\mu\nu}\,f(\kappa r)\,\e^{{\rm i}\phi} \},\ 
\phi = \ell\varphi + \vec{k}\vec{z}-\omega t
,\label{multi.ansatz}\end{equation}
where \mbox{$\vec{k}\vec{z}=\delta_{jk}k^jz^k$}, $f$ is a~function yet to be determined as is the amplitude matrix $A_{\mu\nu}.$
The parameters of the solution are $\omega$, $\kappa$, $\vec{k}$ and $\ell.$

We assume that the $A_{\mu\nu}$ are constants in the Minkowski coordinates and this is the coordinate system we shall work with.

\subsection{\label{sec:general.D.wave.equations}Wave equations}
The wave equations 
for the trace-reversed perturbation \mbox{$\bar{h}_{\mu\nu}=\bar{A}_{\mu\nu}f\e^{{\rm i}\phi}$}
given by
$h_{\mu\nu}=\bar{h}_{\mu\nu}+\frac{1}{2-D}\bar{h}\eta_{\mu\nu}$
yield
\begin{equation}
0 = %\Box\bar{h}_{\mu\nu} =
\left[\omega^2-\vec{k}^2+\frac{f''\kappa^2 + f'\kappa r(B-1)}{r^2f} - \frac{\ell^2}{r^2_2}\right]\bar{h}_{\mu\nu},
\label{multi.radial}\end{equation}
where \mbox{$f'(u) = \frac{\d f}{\d u}$} and \mbox{$r_2=\sqrt{(x^1)^2+(x^2)^2}$}.
The equation can be satisfied only if the last two terms (fractions) add up to a~constant; we shall write it as $-\kappa^2\mathcal{C}$ which has a~correct physical/quantity-dimension.
Of course, $\mathcal{C}$ may depend on both $B$ and $\ell$, i.e., \mbox{$\mathcal{C}=\mathcal{C}(B, \ell)$}.
Consequently, vanishing of the square bracket term in Eq.~\eqref{multi.radial} leads to an equation for the radial function and a~dispersion relation $\omega(\kappa, \vec{k}).$
%$\omega\left(\kappa, \vec{k}\right).$
The equation depends on $x^a$ through $r$ only in the following two subcases.
In the first, \mbox{$B=2$}, and then \mbox{$r~\equiv~r_2$}.
In the second, \mbox{$\ell=0$} and $B$ is arbitrary; then any angular dependence (present in the $r_2$) drops out from the term enclosed in brackets in Eq.~\eqref{multi.radial}.

In these two cases only, we can rewrite the radial-function part of Eq.~\eqref{multi.radial} using a~dimensionless variable \mbox{$u=\kappa r$},
$$f'' + \frac{B-1}{u}f' + \left[\mathcal{C} - \frac{\ell^2}{u^2}\right]f = 0,\; f' = \frac{\d f(u)}{\d u}$$
and one can see that this is a~redressed form of the relation 8.491-3 from~\cite{gradshteyn.ryzhik.1963} (with \mbox{$\alpha=0$} and \mbox{$\beta=\gamma=1$}) leading to 
Bessel functions~$Z_{\nu}$ which can be written as a~linear combination of the Bessel functions of the first and of the second kind for~$\nu\in\mathbb{Z}$.
%\begin{equation}
%f'' + \frac{1-2\alpha}{u}f' + \left[\beta^2\gamma^2u^{2\gamma-2} + \frac{\alpha^2-\nu^2\gamma^2}{u^2}\right]f = 0
%\label{radial.equation.cited}\end{equation}
%solution of which is already known to be~\cite{gradshteyn.ryzhik.1963} (their relation 8.491)
%$$f(u) = u^{\alpha}\left[A_1{\rm J}_{\nu}(\beta u^{\gamma}) + A_2\left\{\begin{array}{rl}
%Y_{\nu}(\beta u^{\gamma}) & \nu\in\mathbb{Z}\\{\rm J}_{-\nu}(\beta u^{\gamma}) & \nu\not\in\mathbb{Z}
%\end{array}\right\}\right],$$
%where ${\rm J}_n$ and $Y_n$ are the Bessel functions of the first and of the second kind, respectively.
%$$f(u) = u^{\alpha}Z_{\nu}(\beta u^{\gamma}),$$
%where $Z_{\nu}$ solves simplified Eq.~\eqref{radial.equation.cited} with \mbox{$\alpha=0$} and \mbox{$\beta=\gamma=1$}.
%Comparing our relation with Eq.~\eqref{radial.equation.cited}, we have \mbox{$\alpha=1-B/2$}, \mbox{$\beta^2 = \mathcal{C}$}, \mbox{$\gamma=1$} and \mbox{$\nu^2=\ell^2+(1-B/2)^2$} so that our solution becomes
The solution is given by
\begin{equation}
%f(u) = u^{1-B/2}\left[A_1{\rm J}_n(\beta u) + A_2\left\{\begin{array}{rl}
%Y_n(\beta u) & n\in\mathbb{Z}\\{\rm J}_{-n}(\beta u) & n\not\in\mathbb{Z}
%\end{array}\right\}\right]
f(u) = u^{1-B/2}Z_{\nu}(\sqrt{\mathcal{C}}\,u),\; \nu^2 = \ell^2 + (1-B/2)^2,
\label{multi.radial.solved}\end{equation}
Since the constant $\mathcal{C}$ is ``arbitrary" as long as Eq.~\eqref{multi.radial} is satisfied, we can set it to be equal to $1$.
This is merely equivalent to rescaling the parameter $\kappa$ in both $f$ and the dispersion relation for $\omega.$

\subsection{\label{sec:general.D.Lorenz.gauge.equations}The Lorenz gauge conditions}
The Lorenz gauge condition \mbox{$0=\bar{h}^{\tau}_{\; \mu, \tau}$} implies
\begin{eqnarray*}
0 %&=& -{\rm i}\omega\bar{A}_{\mu 0} + \bar{A}_{\mu b}\delta^{ab}\left[\ell\varphi_{, a} + \frac{\kappa f'r_{,a}}{f}\right] + {\rm i}\omega\bar{A}_{\mu m}k^m
%\\
&=& {\rm i}\omega\bar{A}_{\mu 0} + {\rm i}\bar{A}_{\mu m}k^m + \sum^{B}_{b=3}\bar{A}_{\mu b}\frac{\kappa f'x^b}{rf}
\\
&& + \left[\bar{A}_{\mu 1}\frac{\kappa f'}{rf}+\bar{A}_{\mu 2}\frac{{\rm i}\ell}{r^2}\right]x^1
+\left[\bar{A}_{\mu 2}\frac{\kappa f'}{rf}-\bar{A}_{\mu 1}\frac{{\rm i}\ell}{r^2}\right]x^2.
\end{eqnarray*}
To analyze the equations, we shall use the parametrization~\eqref{hypersphereparam}.
Let us first deal with the explicitly appearing $x^a$ coordinates.
The last line in the above equations contains only two terms that depend on $\varphi$, $x^1\propto\cos\varphi$ and $x^2\propto\sin\varphi.$
The equations are to be satisfied for all possible values of $\varphi$ and other angular variables, hence both mentioned terms must vanish independently.
Because they vanish, the \mbox{$b=3$} term from the sum is the only one that contains $\vartheta^{B-2}$; thus it must vanish as well, implying \mbox{$\bar{A}_{\mu 3}=0$}.
The process is similar for the \mbox{$b=4$} term and so on for the whole range of the sum.

Let us discuss the vanishing of terms at $x^1$ and $x^2$. 
We split the analysis to the same subcases as in the case of the wave equations.
The first case, \mbox{$B=2$}, gives either that both $\bar{A}_{\mu 1}$ and $\bar{A}_{\mu 2}$ vanish or that
$$\bar{A}_{\mu 2} = \pm{\rm i}\bar{A}_{\mu 1}\neq 0,\; f = K(\kappa r)^{\pm \ell},
%\Rightarrow\mathcal{C}(2, \ell) = 0
$$
where $K$ is a~constant. Equation~\eqref{multi.radial} implies that \mbox{$\mathcal{C} = 0$}.
The second case, \mbox{$\ell=0$}, yields either that both $\bar{A}_{\mu 1}$ and $\bar{A}_{\mu 2}$ vanish or that \mbox{$f'=0$}; i.e., $f$ is a~constant.

The above-mentioned implications for radial functions~$f$ are very restrictive and do not allow us to find a~solution with $f$ being a~Bessel function.
Therefore, we shall ignore these restrictive solutions in the following; i.e. we consider only the case of \mbox{$\bar{A}_{\mu 1}=\bar{A}_{\mu 2}=0$}.

%Now, the remaining $\bar{A}_{\mu 0}$ and $\bar{A}_{\mu m}$ components.
There is a~natural way to construct the $\bar{A}_{0m}$ and $\bar{A}_{nm}$ components using the vector $\vec{k}$ as follows:
\begin{equation}
\bar{A}_{0m} = \tilde{c}\frac{k_m}{k^2},\;
\bar{A}_{nm} = \tilde{a}\frac{k_nk_m}{k^2} + \tilde{b}\left[\frac{k_nk_m}{k^2} - \delta_{nm}\right]
,\label{amplitudes.k.components}\end{equation}
i.e., projecting along the vector $\vec{k}$ (terms with $\tilde{c}$ and $\tilde{a}$) and perpendicular to it (terms with $\tilde{b}$).
Continuing the analysis with the ansatz~\eqref{amplitudes.k.components}, we obtain 
$\bar{A}_{00}=-\tilde{c}/\omega$ and \mbox{$\tilde{a}=-\tilde{c}\omega/k^2$}.
%$$\bar{A}_{00} = \frac{\tilde{c}}{\omega},\; \tilde{a} = \frac{\tilde{c}\omega}{k^2}.$$

Thus we have found a~solution for the amplitudes $\bar{A}_{\mu\nu}$ of the metric perturbation $\bar{h}_{\mu\nu}$.
We need to transform back to $h_{\mu\nu}$ by subtracting the appropriate multiple of the trace.
The result is presented in Sec.~\ref{sec:calculations}, Eq.~\eqref{multi.amplitude}.

\bibliography{gvw_arxiv}

\providecommand{\noopsort}[1]{}\providecommand{\singleletter}[1]{#1}%
\begin{thebibliography}{61}%
\makeatletter
\providecommand \@ifxundefined [1]{%
 \@ifx{#1\undefined}
}%
\providecommand \@ifnum [1]{%
 \ifnum #1\expandafter \@firstoftwo
 \else \expandafter \@secondoftwo
 \fi
}%
\providecommand \@ifx [1]{%
 \ifx #1\expandafter \@firstoftwo
 \else \expandafter \@secondoftwo
 \fi
}%
\providecommand \natexlab [1]{#1}%
\providecommand \enquote  [1]{``#1''}%
\providecommand \bibnamefont  [1]{#1}%
\providecommand \bibfnamefont [1]{#1}%
\providecommand \citenamefont [1]{#1}%
\providecommand \href@noop [0]{\@secondoftwo}%
\providecommand \href [0]{\begingroup \@sanitize@url \@href}%
\providecommand \@href[1]{\@@startlink{#1}\@@href}%
\providecommand \@@href[1]{\endgroup#1\@@endlink}%
\providecommand \@sanitize@url [0]{\catcode `\\12\catcode `\$12\catcode
  `\&12\catcode `\#12\catcode `\^12\catcode `\_12\catcode `\%12\relax}%
\providecommand \@@startlink[1]{}%
\providecommand \@@endlink[0]{}%
\providecommand \url  [0]{\begingroup\@sanitize@url \@url }%
\providecommand \@url [1]{\endgroup\@href {#1}{\urlprefix }}%
\providecommand \urlprefix  [0]{URL }%
\providecommand \Eprint [0]{\href }%
\providecommand \doibase [0]{http://dx.doi.org/}%
\providecommand \selectlanguage [0]{\@gobble}%
\providecommand \bibinfo  [0]{\@secondoftwo}%
\providecommand \bibfield  [0]{\@secondoftwo}%
\providecommand \translation [1]{[#1]}%
\providecommand \BibitemOpen [0]{}%
\providecommand \bibitemStop [0]{}%
\providecommand \bibitemNoStop [0]{.\EOS\space}%
\providecommand \EOS [0]{\spacefactor3000\relax}%
\providecommand \BibitemShut  [1]{\csname bibitem#1\endcsname}%
\let\auto@bib@innerbib\@empty
%</preamble>
\bibitem [{\citenamefont {Abbott}\ \emph {et~al.}(2016)\citenamefont {Abbott}
  \emph {et~al.}}]{ligo}%
  \BibitemOpen
  \bibfield  {author} {\bibinfo {author} {\bibfnamefont {B.~P.}\ \bibnamefont
  {Abbott}} \emph {et~al.} (\bibinfo {collaboration} {LIGO Scientific
  Collaboration and Virgo Collaboration}),\ }\href@noop {} {\bibfield
  {journal} {\bibinfo  {journal} {Phys. Rev. Lett.}\ }\textbf {\bibinfo
  {volume} {116}},\ \bibinfo {pages} {061102} (\bibinfo {year}
  {2016})}\BibitemShut {NoStop}%
\bibitem [{\citenamefont {Abbott}\ \emph {et~al.}(2017)\citenamefont {Abbott}
  \emph {et~al.}}]{ligo2}%
  \BibitemOpen
  \bibfield  {author} {\bibinfo {author} {\bibfnamefont {B.~P.}\ \bibnamefont
  {Abbott}} \emph {et~al.} (\bibinfo {collaboration} {LIGO Scientific
  Collaboration and Virgo Collaboration}),\ }\href@noop {} {\bibfield
  {journal} {\bibinfo  {journal} {Phys. Rev. Lett.}\ }\textbf {\bibinfo
  {volume} {119}},\ \bibinfo {pages} {161101} (\bibinfo {year}
  {2017})}\BibitemShut {NoStop}%
\bibitem [{\citenamefont {Pardo}\ \emph {et~al.}(2018)\citenamefont {Pardo},
  \citenamefont {Fishbach}, \citenamefont {Holz},\ and\ \citenamefont
  {Spergel}}]{pardo}%
  \BibitemOpen
  \bibfield  {author} {\bibinfo {author} {\bibfnamefont {K.}~\bibnamefont
  {Pardo}}, \bibinfo {author} {\bibfnamefont {M.}~\bibnamefont {Fishbach}},
  \bibinfo {author} {\bibfnamefont {D.~E.}\ \bibnamefont {Holz}}, \ and\
  \bibinfo {author} {\bibfnamefont {D.~N.}\ \bibnamefont {Spergel}},\
  }\href@noop {} {\bibfield  {journal} {\bibinfo  {journal} {J. Cosmol.
  Astropart. Phys.}\ }\textbf {\bibinfo {volume} {2018}},\ \bibinfo {pages}
  {048} (\bibinfo {year} {2018})}\BibitemShut {NoStop}%
\bibitem [{\citenamefont {Barnett}(2014)}]{barnett}%
  \BibitemOpen
  \bibfield  {author} {\bibinfo {author} {\bibfnamefont {S.~M.}\ \bibnamefont
  {Barnett}},\ }\href {http://stacks.iop.org/1367-2630/16/i=2/a=023027}
  {\bibfield  {journal} {\bibinfo  {journal} {New J. Phys.}\ }\textbf {\bibinfo
  {volume} {16}},\ \bibinfo {pages} {023027} (\bibinfo {year}
  {2014})}\BibitemShut {NoStop}%
\bibitem [{\citenamefont {Owen}\ \emph {et~al.}(2011)\citenamefont {Owen},
  \citenamefont {Brink}, \citenamefont {Chen}, \citenamefont {Kaplan},
  \citenamefont {Lovelace}, \citenamefont {Matthews}, \citenamefont {Nichols},
  \citenamefont {Scheel}, \citenamefont {Zhang}, \citenamefont {Zimmerman},\
  and\ \citenamefont {Thorne}}]{owen}%
  \BibitemOpen
  \bibfield  {author} {\bibinfo {author} {\bibfnamefont {R.}~\bibnamefont
  {Owen}}, \bibinfo {author} {\bibfnamefont {J.}~\bibnamefont {Brink}},
  \bibinfo {author} {\bibfnamefont {Y.}~\bibnamefont {Chen}}, \bibinfo {author}
  {\bibfnamefont {J.~D.}\ \bibnamefont {Kaplan}}, \bibinfo {author}
  {\bibfnamefont {G.}~\bibnamefont {Lovelace}}, \bibinfo {author}
  {\bibfnamefont {K.~D.}\ \bibnamefont {Matthews}}, \bibinfo {author}
  {\bibfnamefont {D.~A.}\ \bibnamefont {Nichols}}, \bibinfo {author}
  {\bibfnamefont {M.~A.}\ \bibnamefont {Scheel}}, \bibinfo {author}
  {\bibfnamefont {F.}~\bibnamefont {Zhang}}, \bibinfo {author} {\bibfnamefont
  {A.}~\bibnamefont {Zimmerman}}, \ and\ \bibinfo {author} {\bibfnamefont
  {K.~S.}\ \bibnamefont {Thorne}},\ }\href@noop {} {\bibfield  {journal}
  {\bibinfo  {journal} {Phys. Rev. Lett.}\ }\textbf {\bibinfo {volume} {106}},\
  \bibinfo {pages} {151101} (\bibinfo {year} {2011})}\BibitemShut {NoStop}%
\bibitem [{\citenamefont {Durnin}\ \emph {et~al.}(1987)\citenamefont {Durnin},
  \citenamefont {Miceli},\ and\ \citenamefont {Eberly}}]{DurninPRL87}%
  \BibitemOpen
  \bibfield  {author} {\bibinfo {author} {\bibfnamefont {J.}~\bibnamefont
  {Durnin}}, \bibinfo {author} {\bibfnamefont {J.~J.}\ \bibnamefont {Miceli}},
  \ and\ \bibinfo {author} {\bibfnamefont {J.}~\bibnamefont {Eberly}},\
  }\href@noop {} {\bibfield  {journal} {\bibinfo  {journal} {Phys. Rev. Lett.}\
  }\textbf {\bibinfo {volume} {58}},\ \bibinfo {pages} {1499} (\bibinfo {year}
  {1987})}\BibitemShut {NoStop}%
\bibitem [{\citenamefont {Grillo}\ \emph
  {et~al.}(2014{\natexlab{a}})\citenamefont {Grillo}, \citenamefont {Karimi},
  \citenamefont {Gazzadi}, \citenamefont {Frabboni}, \citenamefont {Dennis},\
  and\ \citenamefont {Boyd}}]{Grillo_PhysRevX_2014}%
  \BibitemOpen
  \bibfield  {author} {\bibinfo {author} {\bibfnamefont {V.}~\bibnamefont
  {Grillo}}, \bibinfo {author} {\bibfnamefont {E.}~\bibnamefont {Karimi}},
  \bibinfo {author} {\bibfnamefont {G.~C.}\ \bibnamefont {Gazzadi}}, \bibinfo
  {author} {\bibfnamefont {S.}~\bibnamefont {Frabboni}}, \bibinfo {author}
  {\bibfnamefont {M.~R.}\ \bibnamefont {Dennis}}, \ and\ \bibinfo {author}
  {\bibfnamefont {R.~W.}\ \bibnamefont {Boyd}},\ }\href@noop {} {\bibfield
  {journal} {\bibinfo  {journal} {Phys. Rev. X}\ }\textbf {\bibinfo {volume}
  {4}},\ \bibinfo {pages} {011013} (\bibinfo {year}
  {2014}{\natexlab{a}})}\BibitemShut {NoStop}%
\bibitem [{\citenamefont {Indebetouw}(1989)}]{IndebetouwJOSAA89}%
  \BibitemOpen
  \bibfield  {author} {\bibinfo {author} {\bibfnamefont {G.}~\bibnamefont
  {Indebetouw}},\ }\href@noop {} {\bibfield  {journal} {\bibinfo  {journal} {J.
  Opt. Soc. Am. A}\ }\textbf {\bibinfo {volume} {6}},\ \bibinfo {pages} {150}
  (\bibinfo {year} {1989})}\BibitemShut {NoStop}%
\bibitem [{\citenamefont {Lapointe}(1992)}]{LapointeOPTLASTECH92}%
  \BibitemOpen
  \bibfield  {author} {\bibinfo {author} {\bibfnamefont {M.~R.}\ \bibnamefont
  {Lapointe}},\ }\href@noop {} {\bibfield  {journal} {\bibinfo  {journal} {Opt.
  Laser Technol.}\ }\textbf {\bibinfo {volume} {24}},\ \bibinfo {pages} {315}
  (\bibinfo {year} {1992})}\BibitemShut {NoStop}%
\bibitem [{\citenamefont {Mc{G}loin}\ and\ \citenamefont
  {Dholakia}(2005)}]{McGloinCP05}%
  \BibitemOpen
  \bibfield  {author} {\bibinfo {author} {\bibfnamefont {D.}~\bibnamefont
  {Mc{G}loin}}\ and\ \bibinfo {author} {\bibfnamefont {K.}~\bibnamefont
  {Dholakia}},\ }\href@noop {} {\bibfield  {journal} {\bibinfo  {journal}
  {Contemp. Phys.}\ }\textbf {\bibinfo {volume} {46}},\ \bibinfo {pages} {15}
  (\bibinfo {year} {2005})}\BibitemShut {NoStop}%
\bibitem [{\citenamefont {Brzobohat{\'y}}\ \emph {et~al.}(2008)\citenamefont
  {Brzobohat{\'y}}, \citenamefont {\v{C}i\v{z}m{\'a}r},\ and\ \citenamefont
  {Zem{\'a}nek}}]{BrzobohatyOE08}%
  \BibitemOpen
  \bibfield  {author} {\bibinfo {author} {\bibfnamefont {O.}~\bibnamefont
  {Brzobohat{\'y}}}, \bibinfo {author} {\bibfnamefont {T.}~\bibnamefont
  {\v{C}i\v{z}m{\'a}r}}, \ and\ \bibinfo {author} {\bibfnamefont
  {P.}~\bibnamefont {Zem{\'a}nek}},\ }\href@noop {} {\bibfield  {journal}
  {\bibinfo  {journal} {Opt. Express}\ }\textbf {\bibinfo {volume} {16}},\
  \bibinfo {pages} {12688} (\bibinfo {year} {2008})}\BibitemShut {NoStop}%
\bibitem [{\citenamefont {Mitri}(2008)}]{MITRI20081604}%
  \BibitemOpen
  \bibfield  {author} {\bibinfo {author} {\bibfnamefont {F.}~\bibnamefont
  {Mitri}},\ }\href@noop {} {\bibfield  {journal} {\bibinfo  {journal} {Ann.
  Phys.}\ }\textbf {\bibinfo {volume} {323}},\ \bibinfo {pages} {1604 }
  (\bibinfo {year} {2008})}\BibitemShut {NoStop}%
\bibitem [{\citenamefont {Jim\'{e}nez}\ \emph {et~al.}(2015)\citenamefont
  {Jim\'{e}nez}, \citenamefont {S\'{a}nchez-Morcillo}, \citenamefont
  {Pic\'{o}}, \citenamefont {Garcia-Raffi}, \citenamefont {Romero-Garcia},\
  and\ \citenamefont {Staliunas}}]{JIMENEZ2015245}%
  \BibitemOpen
  \bibfield  {author} {\bibinfo {author} {\bibfnamefont {N.}~\bibnamefont
  {Jim\'{e}nez}}, \bibinfo {author} {\bibfnamefont {V.}~\bibnamefont
  {S\'{a}nchez-Morcillo}}, \bibinfo {author} {\bibfnamefont {R.}~\bibnamefont
  {Pic\'{o}}}, \bibinfo {author} {\bibfnamefont {L.}~\bibnamefont
  {Garcia-Raffi}}, \bibinfo {author} {\bibfnamefont {V.}~\bibnamefont
  {Romero-Garcia}}, \ and\ \bibinfo {author} {\bibfnamefont {K.}~\bibnamefont
  {Staliunas}},\ }\href@noop {} {\bibfield  {journal} {\bibinfo  {journal}
  {Phys. Procedia}\ }\textbf {\bibinfo {volume} {70}},\ \bibinfo {pages} {245 }
  (\bibinfo {year} {2015})},\ \bibinfo {note} {proceedings of the 2015 ICU
  International Congress on Ultrasonics, Metz, France}\BibitemShut {NoStop}%
\bibitem [{\citenamefont {Gorlach}\ \emph {et~al.}(2017)\citenamefont
  {Gorlach}, \citenamefont {Gorlach}, \citenamefont {Lavrinenko},\ and\
  \citenamefont {Novitsky}}]{gorlach}%
  \BibitemOpen
  \bibfield  {author} {\bibinfo {author} {\bibfnamefont {A.~A.}\ \bibnamefont
  {Gorlach}}, \bibinfo {author} {\bibfnamefont {M.~A.}\ \bibnamefont
  {Gorlach}}, \bibinfo {author} {\bibfnamefont {A.~V.}\ \bibnamefont
  {Lavrinenko}}, \ and\ \bibinfo {author} {\bibfnamefont {A.}~\bibnamefont
  {Novitsky}},\ }\href@noop {} {\bibfield  {journal} {\bibinfo  {journal}
  {Phys. Rev. Lett.}\ }\textbf {\bibinfo {volume} {118}},\ \bibinfo {pages}
  {180401} (\bibinfo {year} {2017})}\BibitemShut {NoStop}%
\bibitem [{\citenamefont {Durnin}(1987)}]{DurninJOSAA87}%
  \BibitemOpen
  \bibfield  {author} {\bibinfo {author} {\bibfnamefont {J.}~\bibnamefont
  {Durnin}},\ }\href@noop {} {\bibfield  {journal} {\bibinfo  {journal} {J.
  Opt. Soc. Am. A}\ }\textbf {\bibinfo {volume} {4}},\ \bibinfo {pages} {651}
  (\bibinfo {year} {1987})}\BibitemShut {NoStop}%
\bibitem [{\citenamefont {Bliokh}\ \emph {et~al.}(2007)\citenamefont {Bliokh},
  \citenamefont {Bliokh}, \citenamefont {Savel’ev},\ and\ \citenamefont
  {Nori}}]{bliokh_semiclassical_2007}%
  \BibitemOpen
  \bibfield  {author} {\bibinfo {author} {\bibfnamefont {K.}~\bibnamefont
  {Bliokh}}, \bibinfo {author} {\bibfnamefont {Y.}~\bibnamefont {Bliokh}},
  \bibinfo {author} {\bibfnamefont {S.}~\bibnamefont {Savel’ev}}, \ and\
  \bibinfo {author} {\bibfnamefont {F.}~\bibnamefont {Nori}},\ }\href@noop {}
  {\bibfield  {journal} {\bibinfo  {journal} {Phys. Rev. Lett.}\ }\textbf
  {\bibinfo {volume} {99}},\ \bibinfo {pages} {190404} (\bibinfo {year}
  {2007})}\BibitemShut {NoStop}%
\bibitem [{\citenamefont {Uchida}\ and\ \citenamefont
  {Tonomura}(2010)}]{uchida_generation_2010}%
  \BibitemOpen
  \bibfield  {author} {\bibinfo {author} {\bibfnamefont {M.}~\bibnamefont
  {Uchida}}\ and\ \bibinfo {author} {\bibfnamefont {A.}~\bibnamefont
  {Tonomura}},\ }\href@noop {} {\bibfield  {journal} {\bibinfo  {journal}
  {Nature}\ }\textbf {\bibinfo {volume} {464}},\ \bibinfo {pages} {737}
  (\bibinfo {year} {2010})}\BibitemShut {NoStop}%
\bibitem [{\citenamefont {Verbeeck}\ \emph {et~al.}(2010)\citenamefont
  {Verbeeck}, \citenamefont {Tian},\ and\ \citenamefont
  {Schattschneider}}]{verbeeck_production_2010}%
  \BibitemOpen
  \bibfield  {author} {\bibinfo {author} {\bibfnamefont {J.}~\bibnamefont
  {Verbeeck}}, \bibinfo {author} {\bibfnamefont {H.}~\bibnamefont {Tian}}, \
  and\ \bibinfo {author} {\bibfnamefont {P.}~\bibnamefont {Schattschneider}},\
  }\href@noop {} {\bibfield  {journal} {\bibinfo  {journal} {Nature}\ }\textbf
  {\bibinfo {volume} {467}} (\bibinfo {year} {2010})}\BibitemShut {NoStop}%
\bibitem [{\citenamefont {McMorran}\ \emph {et~al.}(2011)\citenamefont
  {McMorran}, \citenamefont {Agrawal}, \citenamefont {Anderson}, \citenamefont
  {Herzing}, \citenamefont {Lezec}, \citenamefont {McClelland},\ and\
  \citenamefont {Unguris}}]{mcmorran_electron_2011}%
  \BibitemOpen
  \bibfield  {author} {\bibinfo {author} {\bibfnamefont {B.~J.}\ \bibnamefont
  {McMorran}}, \bibinfo {author} {\bibfnamefont {A.}~\bibnamefont {Agrawal}},
  \bibinfo {author} {\bibfnamefont {I.~M.}\ \bibnamefont {Anderson}}, \bibinfo
  {author} {\bibfnamefont {A.~A.}\ \bibnamefont {Herzing}}, \bibinfo {author}
  {\bibfnamefont {H.~J.}\ \bibnamefont {Lezec}}, \bibinfo {author}
  {\bibfnamefont {J.~J.}\ \bibnamefont {McClelland}}, \ and\ \bibinfo {author}
  {\bibfnamefont {J.}~\bibnamefont {Unguris}},\ }\href@noop {} {\bibfield
  {journal} {\bibinfo  {journal} {Science}\ }\textbf {\bibinfo {volume}
  {331}},\ \bibinfo {pages} {192} (\bibinfo {year} {2011})}\BibitemShut
  {NoStop}%
\bibitem [{\citenamefont {Grillo}\ \emph
  {et~al.}(2014{\natexlab{b}})\citenamefont {Grillo}, \citenamefont
  {Carlo~Gazzadi}, \citenamefont {Karimi}, \citenamefont {Mafakheri},
  \citenamefont {Boyd},\ and\ \citenamefont {Frabboni}}]{grillo_highly_2014}%
  \BibitemOpen
  \bibfield  {author} {\bibinfo {author} {\bibfnamefont {V.}~\bibnamefont
  {Grillo}}, \bibinfo {author} {\bibfnamefont {G.}~\bibnamefont
  {Carlo~Gazzadi}}, \bibinfo {author} {\bibfnamefont {E.}~\bibnamefont
  {Karimi}}, \bibinfo {author} {\bibfnamefont {E.}~\bibnamefont {Mafakheri}},
  \bibinfo {author} {\bibfnamefont {R.~W.}\ \bibnamefont {Boyd}}, \ and\
  \bibinfo {author} {\bibfnamefont {S.}~\bibnamefont {Frabboni}},\ }\href
  {http://scitation.aip.org/content/aip/journal/apl/104/4/10.1063/1.4863564}
  {\bibfield  {journal} {\bibinfo  {journal} {Appl. Phys. Lett.}\ }\textbf
  {\bibinfo {volume} {104}},\ \bibinfo {pages} {043109} (\bibinfo {year}
  {2014}{\natexlab{b}})}\BibitemShut {NoStop}%
\bibitem [{\citenamefont {Lloyd}\ \emph {et~al.}(2012)\citenamefont {Lloyd},
  \citenamefont {Babiker},\ and\ \citenamefont {Yuan}}]{lloyd_quantized_2012}%
  \BibitemOpen
  \bibfield  {author} {\bibinfo {author} {\bibfnamefont {S.}~\bibnamefont
  {Lloyd}}, \bibinfo {author} {\bibfnamefont {M.}~\bibnamefont {Babiker}}, \
  and\ \bibinfo {author} {\bibfnamefont {J.}~\bibnamefont {Yuan}},\ }\href@noop
  {} {\bibfield  {journal} {\bibinfo  {journal} {Phys. Rev. Lett.}\ }\textbf
  {\bibinfo {volume} {108}} (\bibinfo {year} {2012})}\BibitemShut {NoStop}%
\bibitem [{\citenamefont {Schattschneider}\ \emph {et~al.}(2014)\citenamefont
  {Schattschneider}, \citenamefont {L{\"o}ffler}, \citenamefont
  {St{\"o}ger-Pollach},\ and\ \citenamefont
  {Verbeeck}}]{schattschneider_is_2014}%
  \BibitemOpen
  \bibfield  {author} {\bibinfo {author} {\bibfnamefont {P.}~\bibnamefont
  {Schattschneider}}, \bibinfo {author} {\bibfnamefont {S.}~\bibnamefont
  {L{\"o}ffler}}, \bibinfo {author} {\bibfnamefont {M.}~\bibnamefont
  {St{\"o}ger-Pollach}}, \ and\ \bibinfo {author} {\bibfnamefont
  {J.}~\bibnamefont {Verbeeck}},\ }\href
  {http://linkinghub.elsevier.com/retrieve/pii/S0304399113001988} {\bibfield
  {journal} {\bibinfo  {journal} {Ultramicroscopy}\ }\textbf {\bibinfo {volume}
  {136}} (\bibinfo {year} {2014})}\BibitemShut {NoStop}%
\bibitem [{\citenamefont {Schachinger}\ \emph {et~al.}(2017)\citenamefont
  {Schachinger}, \citenamefont {L{\"o}ffler}, \citenamefont
  {Steiger-Thirsfeld}, \citenamefont {St{\"o}ger-Pollach}, \citenamefont
  {Schneider}, \citenamefont {Pohl}, \citenamefont {Rellinghaus},\ and\
  \citenamefont {Schattschneider}}]{schachinger_emcd_2017}%
  \BibitemOpen
  \bibfield  {author} {\bibinfo {author} {\bibfnamefont {T.}~\bibnamefont
  {Schachinger}}, \bibinfo {author} {\bibfnamefont {S.}~\bibnamefont
  {L{\"o}ffler}}, \bibinfo {author} {\bibfnamefont {A.}~\bibnamefont
  {Steiger-Thirsfeld}}, \bibinfo {author} {\bibfnamefont {M.}~\bibnamefont
  {St{\"o}ger-Pollach}}, \bibinfo {author} {\bibfnamefont {S.}~\bibnamefont
  {Schneider}}, \bibinfo {author} {\bibfnamefont {D.}~\bibnamefont {Pohl}},
  \bibinfo {author} {\bibfnamefont {B.}~\bibnamefont {Rellinghaus}}, \ and\
  \bibinfo {author} {\bibfnamefont {P.}~\bibnamefont {Schattschneider}},\
  }\href@noop {} {\bibfield  {journal} {\bibinfo  {journal} {Ultramicroscopy}\
  }\textbf {\bibinfo {volume} {179}} (\bibinfo {year} {2017})}\BibitemShut
  {NoStop}%
\bibitem [{\citenamefont {Ugarte}\ and\ \citenamefont
  {Ducati}(2016)}]{ugarte_controlling_2016}%
  \BibitemOpen
  \bibfield  {author} {\bibinfo {author} {\bibfnamefont {D.}~\bibnamefont
  {Ugarte}}\ and\ \bibinfo {author} {\bibfnamefont {C.}~\bibnamefont
  {Ducati}},\ }\href@noop {} {\bibfield  {journal} {\bibinfo  {journal} {Phys.
  Rev. B}\ }\textbf {\bibinfo {volume} {93}} (\bibinfo {year}
  {2016})}\BibitemShut {NoStop}%
\bibitem [{\citenamefont {Guzzinati}\ \emph {et~al.}(2017)\citenamefont
  {Guzzinati}, \citenamefont {B\'{e}ch\'{e}}, \citenamefont
  {Lourenço-Martins}, \citenamefont {Martin}, \citenamefont {Kociak},\ and\
  \citenamefont {Verbeeck}}]{guzzinati_probing_2017}%
  \BibitemOpen
  \bibfield  {author} {\bibinfo {author} {\bibfnamefont {G.}~\bibnamefont
  {Guzzinati}}, \bibinfo {author} {\bibfnamefont {A.}~\bibnamefont
  {B\'{e}ch\'{e}}}, \bibinfo {author} {\bibfnamefont {H.}~\bibnamefont
  {Lourenço-Martins}}, \bibinfo {author} {\bibfnamefont {J.}~\bibnamefont
  {Martin}}, \bibinfo {author} {\bibfnamefont {M.}~\bibnamefont {Kociak}}, \
  and\ \bibinfo {author} {\bibfnamefont {J.}~\bibnamefont {Verbeeck}},\
  }\href@noop {} {\bibfield  {journal} {\bibinfo  {journal} {Nat. Commun.}\
  }\textbf {\bibinfo {volume} {8}},\ \bibinfo {pages} {14999} (\bibinfo {year}
  {2017})}\BibitemShut {NoStop}%
\bibitem [{\citenamefont {Juchtmans}\ \emph {et~al.}(2015)\citenamefont
  {Juchtmans}, \citenamefont {B\'{e}ch\'{e}}, \citenamefont {Abakumov},
  \citenamefont {Batuk},\ and\ \citenamefont
  {Verbeeck}}]{juchtmans_using_2015}%
  \BibitemOpen
  \bibfield  {author} {\bibinfo {author} {\bibfnamefont {R.}~\bibnamefont
  {Juchtmans}}, \bibinfo {author} {\bibfnamefont {A.}~\bibnamefont
  {B\'{e}ch\'{e}}}, \bibinfo {author} {\bibfnamefont {A.}~\bibnamefont
  {Abakumov}}, \bibinfo {author} {\bibfnamefont {M.}~\bibnamefont {Batuk}}, \
  and\ \bibinfo {author} {\bibfnamefont {J.}~\bibnamefont {Verbeeck}},\
  }\href@noop {} {\bibfield  {journal} {\bibinfo  {journal} {Phys. Rev. B}\
  }\textbf {\bibinfo {volume} {91}},\ \bibinfo {pages} {094112} (\bibinfo
  {year} {2015})}\BibitemShut {NoStop}%
\bibitem [{\citenamefont {Verbeeck}\ \emph {et~al.}(2013)\citenamefont
  {Verbeeck}, \citenamefont {Tian},\ and\ \citenamefont
  {Van~Tendeloo}}]{verbeeck_how_2013}%
  \BibitemOpen
  \bibfield  {author} {\bibinfo {author} {\bibfnamefont {J.}~\bibnamefont
  {Verbeeck}}, \bibinfo {author} {\bibfnamefont {H.}~\bibnamefont {Tian}}, \
  and\ \bibinfo {author} {\bibfnamefont {G.}~\bibnamefont {Van~Tendeloo}},\
  }\href@noop {} {\bibfield  {journal} {\bibinfo  {journal} {Adv. Mater.}\
  }\textbf {\bibinfo {volume} {25}},\ \bibinfo {pages} {1114} (\bibinfo {year}
  {2013})}\BibitemShut {NoStop}%
\bibitem [{\citenamefont {Karimi}\ \emph {et~al.}(2012)\citenamefont {Karimi},
  \citenamefont {Marrucci}, \citenamefont {Grillo},\ and\ \citenamefont
  {Santamato}}]{karimi_spin--orbital_2012}%
  \BibitemOpen
  \bibfield  {author} {\bibinfo {author} {\bibfnamefont {E.}~\bibnamefont
  {Karimi}}, \bibinfo {author} {\bibfnamefont {L.}~\bibnamefont {Marrucci}},
  \bibinfo {author} {\bibfnamefont {V.}~\bibnamefont {Grillo}}, \ and\ \bibinfo
  {author} {\bibfnamefont {E.}~\bibnamefont {Santamato}},\ }\href@noop {}
  {\bibfield  {journal} {\bibinfo  {journal} {Phys. Rev. Lett.}\ }\textbf
  {\bibinfo {volume} {108}} (\bibinfo {year} {2012})}\BibitemShut {NoStop}%
\bibitem [{\citenamefont {Einstein}\ and\ \citenamefont
  {Rosen}(1937)}]{einstein1937}%
  \BibitemOpen
  \bibfield  {author} {\bibinfo {author} {\bibfnamefont {A.}~\bibnamefont
  {Einstein}}\ and\ \bibinfo {author} {\bibfnamefont {N.}~\bibnamefont
  {Rosen}},\ }\href@noop {} {\bibfield  {journal} {\bibinfo  {journal} {J.
  Franklin Inst.}\ }\textbf {\bibinfo {volume} {223}},\ \bibinfo {pages} {43}
  (\bibinfo {year} {1937})}\BibitemShut {NoStop}%
\bibitem [{\citenamefont {Weber}(2013)}]{weber}%
  \BibitemOpen
  \bibfield  {author} {\bibinfo {author} {\bibfnamefont {J.}~\bibnamefont
  {Weber}},\ }\href {https://books.google.cz/books?id=lgDDAgAAQBAJ} {\emph
  {\bibinfo {title} {General Relativity and Gravitational Waves}}}\ (\bibinfo
  {publisher} {Dover Publications},\ \bibinfo {year} {2013})\BibitemShut
  {NoStop}%
\bibitem [{\citenamefont {Misner}\ \emph {et~al.}(1973)\citenamefont {Misner},
  \citenamefont {Thorne},\ and\ \citenamefont {Wheeler}}]{mtw}%
  \BibitemOpen
  \bibfield  {author} {\bibinfo {author} {\bibfnamefont {C.}~\bibnamefont
  {Misner}}, \bibinfo {author} {\bibfnamefont {K.}~\bibnamefont {Thorne}}, \
  and\ \bibinfo {author} {\bibfnamefont {J.}~\bibnamefont {Wheeler}},\ }\href
  {https://books.google.cz/books?id=w4Gigq3tY1kC} {\emph {\bibinfo {title}
  {Gravitation}}}\ (\bibinfo  {publisher} {W. H. Freeman},\ \bibinfo {year}
  {1973})\BibitemShut {NoStop}%
\bibitem [{\citenamefont {Chishtie}\ \emph {et~al.}(2005)\citenamefont
  {Chishtie}, \citenamefont {Valluri}, \citenamefont {Rao}, \citenamefont
  {Sikorski},\ and\ \citenamefont {Williams}}]{chishtie}%
  \BibitemOpen
  \bibfield  {author} {\bibinfo {author} {\bibfnamefont {F.}~\bibnamefont
  {Chishtie}}, \bibinfo {author} {\bibfnamefont {S.}~\bibnamefont {Valluri}},
  \bibinfo {author} {\bibfnamefont {K.}~\bibnamefont {Rao}}, \bibinfo {author}
  {\bibfnamefont {D.}~\bibnamefont {Sikorski}}, \ and\ \bibinfo {author}
  {\bibfnamefont {T.}~\bibnamefont {Williams}},\ }\href@noop {} {\bibfield
  {journal} {\bibinfo  {journal} {Proceedings of the 19th International
  Symposium on High Performance Computing Systems and Applications, Univ Guelp,
  Guelph, Canada}\ ,\ \bibinfo {pages} {34}} (\bibinfo {year}
  {2005})}\BibitemShut {NoStop}%
\bibitem [{\citenamefont {Nichols}\ \emph {et~al.}(2011)\citenamefont
  {Nichols}, \citenamefont {Owen}, \citenamefont {Zhang}, \citenamefont
  {Zimmerman}, \citenamefont {Brink}, \citenamefont {Chen}, \citenamefont
  {Kaplan}, \citenamefont {Lovelace}, \citenamefont {Matthews}, \citenamefont
  {Scheel},\ and\ \citenamefont {Thorne}}]{nichols}%
  \BibitemOpen
  \bibfield  {author} {\bibinfo {author} {\bibfnamefont {D.~A.}\ \bibnamefont
  {Nichols}}, \bibinfo {author} {\bibfnamefont {R.}~\bibnamefont {Owen}},
  \bibinfo {author} {\bibfnamefont {F.}~\bibnamefont {Zhang}}, \bibinfo
  {author} {\bibfnamefont {A.}~\bibnamefont {Zimmerman}}, \bibinfo {author}
  {\bibfnamefont {J.}~\bibnamefont {Brink}}, \bibinfo {author} {\bibfnamefont
  {Y.}~\bibnamefont {Chen}}, \bibinfo {author} {\bibfnamefont {J.~D.}\
  \bibnamefont {Kaplan}}, \bibinfo {author} {\bibfnamefont {G.}~\bibnamefont
  {Lovelace}}, \bibinfo {author} {\bibfnamefont {K.~D.}\ \bibnamefont
  {Matthews}}, \bibinfo {author} {\bibfnamefont {M.~A.}\ \bibnamefont
  {Scheel}}, \ and\ \bibinfo {author} {\bibfnamefont {K.~S.}\ \bibnamefont
  {Thorne}},\ }\href@noop {} {\bibfield  {journal} {\bibinfo  {journal} {Phys.
  Rev. D}\ }\textbf {\bibinfo {volume} {84}},\ \bibinfo {pages} {124014}
  (\bibinfo {year} {2011})}\BibitemShut {NoStop}%
\bibitem [{\citenamefont {Zhang}\ \emph {et~al.}(2012)\citenamefont {Zhang},
  \citenamefont {Zimmerman}, \citenamefont {Nichols}, \citenamefont {Chen},
  \citenamefont {Lovelace}, \citenamefont {Matthews}, \citenamefont {Owen},\
  and\ \citenamefont {Thorne}}]{zhang}%
  \BibitemOpen
  \bibfield  {author} {\bibinfo {author} {\bibfnamefont {F.}~\bibnamefont
  {Zhang}}, \bibinfo {author} {\bibfnamefont {A.}~\bibnamefont {Zimmerman}},
  \bibinfo {author} {\bibfnamefont {D.~A.}\ \bibnamefont {Nichols}}, \bibinfo
  {author} {\bibfnamefont {Y.}~\bibnamefont {Chen}}, \bibinfo {author}
  {\bibfnamefont {G.}~\bibnamefont {Lovelace}}, \bibinfo {author}
  {\bibfnamefont {K.~D.}\ \bibnamefont {Matthews}}, \bibinfo {author}
  {\bibfnamefont {R.}~\bibnamefont {Owen}}, \ and\ \bibinfo {author}
  {\bibfnamefont {K.~S.}\ \bibnamefont {Thorne}},\ }\href@noop {} {\bibfield
  {journal} {\bibinfo  {journal} {Phys. Rev. D}\ }\textbf {\bibinfo {volume}
  {86}},\ \bibinfo {pages} {084049} (\bibinfo {year} {2012})}\BibitemShut
  {NoStop}%
\bibitem [{\citenamefont {Nichols}\ \emph {et~al.}(2012)\citenamefont
  {Nichols}, \citenamefont {Zimmerman}, \citenamefont {Chen}, \citenamefont
  {Lovelace}, \citenamefont {Matthews}, \citenamefont {Owen}, \citenamefont
  {Zhang},\ and\ \citenamefont {Thorne}}]{nichols2}%
  \BibitemOpen
  \bibfield  {author} {\bibinfo {author} {\bibfnamefont {D.~A.}\ \bibnamefont
  {Nichols}}, \bibinfo {author} {\bibfnamefont {A.}~\bibnamefont {Zimmerman}},
  \bibinfo {author} {\bibfnamefont {Y.}~\bibnamefont {Chen}}, \bibinfo {author}
  {\bibfnamefont {G.}~\bibnamefont {Lovelace}}, \bibinfo {author}
  {\bibfnamefont {K.~D.}\ \bibnamefont {Matthews}}, \bibinfo {author}
  {\bibfnamefont {R.}~\bibnamefont {Owen}}, \bibinfo {author} {\bibfnamefont
  {F.}~\bibnamefont {Zhang}}, \ and\ \bibinfo {author} {\bibfnamefont {K.~S.}\
  \bibnamefont {Thorne}},\ }\href@noop {} {\bibfield  {journal} {\bibinfo
  {journal} {Phys. Rev. D}\ }\textbf {\bibinfo {volume} {86}},\ \bibinfo
  {pages} {104028} (\bibinfo {year} {2012})}\BibitemShut {NoStop}%
\bibitem [{\citenamefont {Zimmerman}\ \emph {et~al.}(2011)\citenamefont
  {Zimmerman}, \citenamefont {Nichols},\ and\ \citenamefont
  {Zhang}}]{zimmerman}%
  \BibitemOpen
  \bibfield  {author} {\bibinfo {author} {\bibfnamefont {A.}~\bibnamefont
  {Zimmerman}}, \bibinfo {author} {\bibfnamefont {D.~A.}\ \bibnamefont
  {Nichols}}, \ and\ \bibinfo {author} {\bibfnamefont {F.}~\bibnamefont
  {Zhang}},\ }\href@noop {} {\bibfield  {journal} {\bibinfo  {journal} {Phys.
  Rev. D}\ }\textbf {\bibinfo {volume} {84}},\ \bibinfo {pages} {044037}
  (\bibinfo {year} {2011})}\BibitemShut {NoStop}%
\bibitem [{\citenamefont {Bialynicki-Birula}\ and\ \citenamefont
  {Bialynicka-Birula}(2016)}]{birula}%
  \BibitemOpen
  \bibfield  {author} {\bibinfo {author} {\bibfnamefont {I.}~\bibnamefont
  {Bialynicki-Birula}}\ and\ \bibinfo {author} {\bibfnamefont {Z.}~\bibnamefont
  {Bialynicka-Birula}},\ }\href
  {http://stacks.iop.org/1367-2630/18/i=2/a=023022} {\bibfield  {journal}
  {\bibinfo  {journal} {New J. Phys.}\ }\textbf {\bibinfo {volume} {18}},\
  \bibinfo {pages} {023022} (\bibinfo {year} {2016})}\BibitemShut {NoStop}%
\bibitem [{\citenamefont {Penrose}(1965)}]{penrosec}%
  \BibitemOpen
  \bibfield  {author} {\bibinfo {author} {\bibfnamefont {R.}~\bibnamefont
  {Penrose}},\ }\href
  {http://rspa.royalsocietypublishing.org/content/284/1397/159} {\bibfield
  {journal} {\bibinfo  {journal} {Proc. R. Soc. A}\ }\textbf {\bibinfo {volume}
  {284}},\ \bibinfo {pages} {159} (\bibinfo {year} {1965})}\BibitemShut
  {NoStop}%
\bibitem [{\citenamefont {Penrose}\ and\ \citenamefont
  {Rindler}(1984)}]{penroseb}%
  \BibitemOpen
  \bibfield  {author} {\bibinfo {author} {\bibfnamefont {R.}~\bibnamefont
  {Penrose}}\ and\ \bibinfo {author} {\bibfnamefont {W.}~\bibnamefont
  {Rindler}},\ }\href {https://books.google.cz/books?id=C1ovtwAACAAJ} {\emph
  {\bibinfo {title} {Spinors and Space-time}}}\ (\bibinfo  {publisher}
  {Cambridge University Press},\ \bibinfo {year} {1984})\BibitemShut {NoStop}%
\bibitem [{\citenamefont {Ilderton}(2018)}]{ilderton}%
  \BibitemOpen
  \bibfield  {author} {\bibinfo {author} {\bibfnamefont {A.}~\bibnamefont
  {Ilderton}},\ }\href@noop {} {\bibfield  {journal} {\bibinfo  {journal}
  {Phys. Lett. B}\ }\textbf {\bibinfo {volume} {782}},\ \bibinfo {pages} {22}
  (\bibinfo {year} {2018})}\BibitemShut {NoStop}%
\bibitem [{\citenamefont {{van der Walt}}\ \emph {et~al.}(2011)\citenamefont
  {{van der Walt}}, \citenamefont {{Colbert}},\ and\ \citenamefont
  {{Varoquaux}}}]{numpy.article}%
  \BibitemOpen
  \bibfield  {author} {\bibinfo {author} {\bibfnamefont {S.}~\bibnamefont {{van
  der Walt}}}, \bibinfo {author} {\bibfnamefont {S.~C.}\ \bibnamefont
  {{Colbert}}}, \ and\ \bibinfo {author} {\bibfnamefont {G.}~\bibnamefont
  {{Varoquaux}}},\ }\href {\doibase 10.1109/MCSE.2011.37} {\bibfield  {journal}
  {\bibinfo  {journal} {Comput. Sci. Eng.}\ }\textbf {\bibinfo {volume} {13}},\
  \bibinfo {pages} {22} (\bibinfo {year} {2011})}\BibitemShut {NoStop}%
\bibitem [{\citenamefont {Oliphant}(2019)}]{numpy.man}%
  \BibitemOpen
  \bibfield  {author} {\bibinfo {author} {\bibfnamefont {T.}~\bibnamefont
  {Oliphant}},\ }\href {http://www.numpy.org/} {\emph {\bibinfo {title}
  {{NumPy}: A guide to {NumPy}}}}\ (\bibinfo {year} {2006--2019})\BibitemShut
  {NoStop}%
\bibitem [{\citenamefont {Jones}\ \emph {et~al.}(2019)\citenamefont {Jones}
  \emph {et~al.}}]{scipy}%
  \BibitemOpen
  \bibfield  {author} {\bibinfo {author} {\bibfnamefont {E.}~\bibnamefont
  {Jones}} \emph {et~al.},\ }\href {http://www.scipy.org/} {\emph {\bibinfo
  {title} {{SciPy}: Open source scientific tools for {Python}}}}\ (\bibinfo
  {year} {2001--2019})\BibitemShut {NoStop}%
\bibitem [{\citenamefont {Meurer}\ \emph {et~al.}(2017)\citenamefont {Meurer}
  \emph {et~al.}}]{sympy}%
  \BibitemOpen
  \bibfield  {author} {\bibinfo {author} {\bibfnamefont {A.}~\bibnamefont
  {Meurer}} \emph {et~al.},\ }\href {\doibase 10.7717/peerj-cs.103} {\bibfield
  {journal} {\bibinfo  {journal} {PeerJ. Comput. Sci.}\ }\textbf {\bibinfo
  {volume} {3}},\ \bibinfo {pages} {e103} (\bibinfo {year} {2017})}\BibitemShut
  {NoStop}%
\bibitem [{\citenamefont {{Hunter}}(2007)}]{matplotlib}%
  \BibitemOpen
  \bibfield  {author} {\bibinfo {author} {\bibfnamefont {J.~D.}\ \bibnamefont
  {{Hunter}}},\ }\href {\doibase 10.1109/MCSE.2007.55} {\bibfield  {journal}
  {\bibinfo  {journal} {Comput. Sci. Eng.}\ }\textbf {\bibinfo {volume} {9}},\
  \bibinfo {pages} {90} (\bibinfo {year} {2007})}\BibitemShut {NoStop}%
\bibitem [{\citenamefont {Ramachandran}\ and\ \citenamefont
  {Varoquaux}(2011)}]{ramachandran2011mayavi}%
  \BibitemOpen
  \bibfield  {author} {\bibinfo {author} {\bibfnamefont {P.}~\bibnamefont
  {Ramachandran}}\ and\ \bibinfo {author} {\bibfnamefont {G.}~\bibnamefont
  {Varoquaux}},\ }\href@noop {} {\bibfield  {journal} {\bibinfo  {journal}
  {Comput. Sci. Eng.}\ }\textbf {\bibinfo {volume} {13}},\ \bibinfo {pages}
  {40} (\bibinfo {year} {2011})}\BibitemShut {NoStop}%
\bibitem [{\citenamefont {Crowley}\ and\ \citenamefont
  {Grant}(2017)}]{crowley}%
  \BibitemOpen
  \bibfield  {author} {\bibinfo {author} {\bibfnamefont {D.}~\bibnamefont
  {Crowley}}\ and\ \bibinfo {author} {\bibfnamefont {M.}~\bibnamefont
  {Grant}},\ }\href
  {https://linkinghub.elsevier.com/retrieve/pii/S0393044017300955} {\bibfield
  {journal} {\bibinfo  {journal} {J. Geom. Phys.}\ }\textbf {\bibinfo {volume}
  {117}},\ \bibinfo {pages} {187} (\bibinfo {year} {2017})}\BibitemShut
  {NoStop}%
\bibitem [{\citenamefont {Delva}\ and\ \citenamefont
  {Ger\v{s}l}(2017)}]{honza}%
  \BibitemOpen
  \bibfield  {author} {\bibinfo {author} {\bibfnamefont {P.}~\bibnamefont
  {Delva}}\ and\ \bibinfo {author} {\bibfnamefont {J.}~\bibnamefont
  {Ger\v{s}l}},\ }\href@noop {} {\bibfield  {journal} {\bibinfo  {journal}
  {Universe}\ }\textbf {\bibinfo {volume} {3}},\ \bibinfo {pages} {24}
  (\bibinfo {year} {2017})}\BibitemShut {NoStop}%
\bibitem [{\citenamefont {Newman}\ and\ \citenamefont
  {Penrose}(1962)}]{newmanpenrose}%
  \BibitemOpen
  \bibfield  {author} {\bibinfo {author} {\bibfnamefont {E.}~\bibnamefont
  {Newman}}\ and\ \bibinfo {author} {\bibfnamefont {R.}~\bibnamefont
  {Penrose}},\ }\href@noop {} {\bibfield  {journal} {\bibinfo  {journal} {J.
  Math. Phys.}\ }\textbf {\bibinfo {volume} {3}},\ \bibinfo {pages} {566}
  (\bibinfo {year} {1962})}\BibitemShut {NoStop}%
\bibitem [{\citenamefont {de~Jes{\'{u}}s Cabrera-Rosas}\ \emph
  {et~al.}(2016)\citenamefont {de~Jes{\'{u}}s Cabrera-Rosas}, \citenamefont
  {Esp{\'{\i}}ndola-Ramos}, \citenamefont {Ju{\'{a}}rez-Reyes}, \citenamefont
  {Juli{\'{a}}n-Mac{\'{\i}}as}, \citenamefont {Ortega-Vidals}, \citenamefont
  {Silva-Ortigoza}, \citenamefont {Silva-Ortigoza},\ and\ \citenamefont
  {Sosa-S{\'{a}}nchez}}]{ssanchez}%
  \BibitemOpen
  \bibfield  {author} {\bibinfo {author} {\bibfnamefont {O.}~\bibnamefont
  {de~Jes{\'{u}}s Cabrera-Rosas}}, \bibinfo {author} {\bibfnamefont
  {E.}~\bibnamefont {Esp{\'{\i}}ndola-Ramos}}, \bibinfo {author} {\bibfnamefont
  {S.~A.}\ \bibnamefont {Ju{\'{a}}rez-Reyes}}, \bibinfo {author} {\bibfnamefont
  {I.}~\bibnamefont {Juli{\'{a}}n-Mac{\'{\i}}as}}, \bibinfo {author}
  {\bibfnamefont {P.}~\bibnamefont {Ortega-Vidals}}, \bibinfo {author}
  {\bibfnamefont {G.}~\bibnamefont {Silva-Ortigoza}}, \bibinfo {author}
  {\bibfnamefont {R.}~\bibnamefont {Silva-Ortigoza}}, \ and\ \bibinfo {author}
  {\bibfnamefont {C.~T.}\ \bibnamefont {Sosa-S{\'{a}}nchez}},\ }\href@noop {}
  {\bibfield  {journal} {\bibinfo  {journal} {J. Opt.}\ }\textbf {\bibinfo
  {volume} {19}},\ \bibinfo {pages} {015603} (\bibinfo {year}
  {2016})}\BibitemShut {NoStop}%
\bibitem [{\citenamefont {Hayward}(2000)}]{hayward}%
  \BibitemOpen
  \bibfield  {author} {\bibinfo {author} {\bibfnamefont {S.~A.}\ \bibnamefont
  {Hayward}},\ }\href@noop {} {\bibfield  {journal} {\bibinfo  {journal}
  {Class. Quantum Grav.}\ }\textbf {\bibinfo {volume} {17}},\ \bibinfo {pages}
  {1749} (\bibinfo {year} {2000})}\BibitemShut {NoStop}%
\bibitem [{\citenamefont {Weber}\ and\ \citenamefont
  {Wheeler}(1957)}]{weberwheeler}%
  \BibitemOpen
  \bibfield  {author} {\bibinfo {author} {\bibfnamefont {J.}~\bibnamefont
  {Weber}}\ and\ \bibinfo {author} {\bibfnamefont {J.~A.}\ \bibnamefont
  {Wheeler}},\ }\href@noop {} {\bibfield  {journal} {\bibinfo  {journal} {Rev.
  Mod. Phys.}\ }\textbf {\bibinfo {volume} {29}},\ \bibinfo {pages} {509}
  (\bibinfo {year} {1957})}\BibitemShut {NoStop}%
\bibitem [{\citenamefont {Gowdy}\ and\ \citenamefont
  {Edmonds}(2007)}]{gowdyedmonds}%
  \BibitemOpen
  \bibfield  {author} {\bibinfo {author} {\bibfnamefont {R.~H.}\ \bibnamefont
  {Gowdy}}\ and\ \bibinfo {author} {\bibfnamefont {B.~D.}\ \bibnamefont
  {Edmonds}},\ }\href@noop {} {\bibfield  {journal} {\bibinfo  {journal} {Phys.
  Rev. D}\ }\textbf {\bibinfo {volume} {75}} (\bibinfo {year}
  {2007})}\BibitemShut {NoStop}%
\bibitem [{\citenamefont {Bi\v{c}{\'a}k}\ and\ \citenamefont
  {Schmidt}(1989)}]{bicak}%
  \BibitemOpen
  \bibfield  {author} {\bibinfo {author} {\bibfnamefont {J.}~\bibnamefont
  {Bi\v{c}{\'a}k}}\ and\ \bibinfo {author} {\bibfnamefont {B.}~\bibnamefont
  {Schmidt}},\ }\href@noop {} {\bibfield  {journal} {\bibinfo  {journal} {Phys.
  Rev. D}\ }\textbf {\bibinfo {volume} {40}},\ \bibinfo {pages} {1827}
  (\bibinfo {year} {1989})}\BibitemShut {NoStop}%
\bibitem [{\citenamefont {Marder}(1958)}]{marder}%
  \BibitemOpen
  \bibfield  {author} {\bibinfo {author} {\bibfnamefont {L.}~\bibnamefont
  {Marder}},\ }\href@noop {} {\bibfield  {journal} {\bibinfo  {journal} {Proc.
  R. Soc. A}\ }\textbf {\bibinfo {volume} {244}},\ \bibinfo {pages} {524}
  (\bibinfo {year} {1958})}\BibitemShut {NoStop}%
\bibitem [{\citenamefont {Anderson}(2003)}]{anderson}%
  \BibitemOpen
  \bibfield  {author} {\bibinfo {author} {\bibfnamefont {M.~R.}\ \bibnamefont
  {Anderson}},\ }\href@noop {} {\emph {\bibinfo {title} {The mathematical
  theory of cosmic strings: cosmic strings in the wire approximation}}}\
  (\bibinfo  {publisher} {Institute of Physics Pub},\ \bibinfo {year}
  {2003})\BibitemShut {NoStop}%
\bibitem [{\citenamefont {Kibble}(1976)}]{kibble}%
  \BibitemOpen
  \bibfield  {author} {\bibinfo {author} {\bibfnamefont {T.~W.~B.}\
  \bibnamefont {Kibble}},\ }\href
  {http://stacks.iop.org/0305-4470/9/i=8/a=029?key=crossref.fd6da1d908a3bf72b56557edd8d38c37}
  {\bibfield  {journal} {\bibinfo  {journal} {J. Phys. A}\ }\textbf {\bibinfo
  {volume} {9}},\ \bibinfo {pages} {1387} (\bibinfo {year} {1976})}\BibitemShut
  {NoStop}%
\bibitem [{\citenamefont {Wen}\ \emph {et~al.}(2014)\citenamefont {Wen},
  \citenamefont {Li}, \citenamefont {Fang},\ and\ \citenamefont
  {Beckwith}}]{wen}%
  \BibitemOpen
  \bibfield  {author} {\bibinfo {author} {\bibfnamefont {H.}~\bibnamefont
  {Wen}}, \bibinfo {author} {\bibfnamefont {F.}~\bibnamefont {Li}}, \bibinfo
  {author} {\bibfnamefont {Z.}~\bibnamefont {Fang}}, \ and\ \bibinfo {author}
  {\bibfnamefont {A.}~\bibnamefont {Beckwith}},\ }\href
  {http://link.springer.com/10.1140/epjc/s10052-014-2998-9} {\bibfield
  {journal} {\bibinfo  {journal} {Eur. Phys. J. C}\ }\textbf {\bibinfo {volume}
  {74}} (\bibinfo {year} {2014})}\BibitemShut {NoStop}%
\bibitem [{\citenamefont {{\v{C}}erm\'{a}k}\ and\ \citenamefont
  {Zouhar}(2014)}]{cermak.zouhar.2014}%
  \BibitemOpen
  \bibfield  {author} {\bibinfo {author} {\bibfnamefont {M.}~\bibnamefont
  {{\v{C}}erm\'{a}k}}\ and\ \bibinfo {author} {\bibfnamefont {M.}~\bibnamefont
  {Zouhar}},\ }\href@noop {} {\bibfield  {journal} {\bibinfo  {journal} {Phys.
  Rev. D}\ }\textbf {\bibinfo {volume} {89}} (\bibinfo {year}
  {2014})}\BibitemShut {NoStop}%
\bibitem [{\citenamefont {{\v{C}}erm\'{a}k}\ and\ \citenamefont
  {Zouhar}(2012)}]{cermak.zouhar.2012}%
  \BibitemOpen
  \bibfield  {author} {\bibinfo {author} {\bibfnamefont {M.}~\bibnamefont
  {{\v{C}}erm\'{a}k}}\ and\ \bibinfo {author} {\bibfnamefont {M.}~\bibnamefont
  {Zouhar}},\ }\href@noop {} {\bibfield  {journal} {\bibinfo  {journal} {Int.
  J. Theor. Phys.}\ }\textbf {\bibinfo {volume} {51}},\ \bibinfo {pages} {2455}
  (\bibinfo {year} {2012})}\BibitemShut {NoStop}%
\bibitem [{\citenamefont {Gradshteyn}\ and\ \citenamefont
  {Ryzhik}(1963)}]{gradshteyn.ryzhik.1963}%
  \BibitemOpen
  \bibfield  {author} {\bibinfo {author} {\bibfnamefont {I.}~\bibnamefont
  {Gradshteyn}}\ and\ \bibinfo {author} {\bibfnamefont {I.}~\bibnamefont
  {Ryzhik}},\ }\href@noop {} {\emph {\bibinfo {title} {Tables of integrals,
  series and products}}},\ \bibinfo {edition} {4th}\ ed.\ (\bibinfo {address}
  {Moscow},\ \bibinfo {year} {1963})\BibitemShut {NoStop}%
\end{thebibliography}%
\end{document}